\documentclass[a4paper]{article}
\usepackage{eepic} 
\begin{document}

\title{On the correlation functions of the domain wall 
       six vertex model\footnote{Supported by the Australian 
       Research Council (ARC).}} 

\author{Omar Foda$^1$ and Ian Preston$^2$\\ 
       {}\\
       $^1$Department of Mathematics and Statistics,\\ 
       The University of Melbourne,\\ 
       Parkville, Victoria 3010, Australia.\\
       {}\\
       $^2$Magdalen College,\\
       Oxford, OX1 4AU, Oxfordshire, UK} 
       
\maketitle

\begin{abstract}
We propose an (essentially combinatorial) approach to the 
correlation functions of the domain wall six vertex model.

We reproduce the boundary 1-point function determinant 
expression of Bogoliubov, Pronko and Zvonarev, then use 
that as a building block to obtain analogous expressions 
for boundary 2-point functions. 

The latter can be used, at least in principle, to express 
more general boundary (and bulk) correlation functions as 
sums over (products of) determinants.

\end{abstract} 

\newtheorem{theo}{Theorem}      
\newtheorem{co}{Corollary}[theo]   
\newtheorem{de}{Definition}     
\newtheorem{pr}{Proposition} 
\newtheorem{re}{Remark} 
\newtheorem{ex}{Example}
\newtheorem{no}{Notation}
\newtheorem{ca}{Figure}

\def\ll{\left\lgroup}
\def\rr{\right\rgroup}

\newtheorem{definition}{Definition}
\newtheorem{theorem}{Theorem}
\newtheorem{remark}{Remark}
\newtheorem{lemma}{Lemma}
\newtheorem{proposition}{Proposition}
\newtheorem{corollary}{Corollary}
\newtheorem{example}{Example}

\section{Introduction}

Computing off critical correlation functions\footnote{The 
literature on off critical correlation functions is extensive. 
We refer the reader to \cite{Korepin-book} for references 
to the literature up to the early 90's and to a search of 
{\tt http://arXiv.org} for more recent literature.} is 
probably the most challenging open problem currently under 
investigation in exactly solvable lattice
models \cite{Baxter-book}.

The six vertex model, with domain wall boundary conditions
(dwbc's) is an ideal testing ground of possible approaches 
to such computations, particularly if one is interested in 
computations on a finite lattice\footnote{The previous
footnote applies {\it verbatim} to the literature on the 
domain wall six vertex model.}.

The model was first introduced by Korepin \cite{Korepin}, 
who also formulated recursion relations that uniquely 
determine the partition function. Korepin's recursion 
relations were solved by Izergin \cite{Izergin}, who 
obtained a determinant representation for the partition 
function. 

Using the algebraic Bethe {\it ansatz} \cite{Korepin-book}, 
Bogoliubov, Pronko and Zvonarev introduced a definition 
of boundary 1-point functions in the model \cite{BPZ}, 
and computed them in determinant form\footnote{Bogoliubov 
{\it et al} obtained determinant expressions for the 
1-point functions, and also for the boundary spontaneous 
polarization, which is a more difficult problem. Here, 
we discuss only the former.}.

Inspired by Bogoliubov {\it et al}, we propose a definition 
of boundary $n$-point functions, that in the specific case
that they considered, (almost) coincides with theirs (up to 
obvious factors due to the differences in our definitions
of what a boundary 1-point function is).

We use combinatorics, based on the Yang Baxter equation, 
to compute determinant representations for boundary 1-point, 
and 2-point functions, and outline the (almost mechanical) 
extension to (certain) boundary $n$-point and bulk correlation 
functions. In the case of boundary 1-point functions, we 
reproduce the result of Bogoliubov {\it et al} \cite{BPZ}
using elementary manipulations. 

Our derivations are (basically) combinatorial and rely on 
repeated application of graphical operations, hence the 
proliferation of figures in the paper\footnote{Our results 
can be produced using the algebraic Bethe {\it ansatz}. 
However, our method is elementary, and hopefully there is 
virtue in having more than one approach to this problem.}. 

There has been interest in this very model from algebraic 
combinatorialists over the past decade, particularly since 
the work of Kuperberg \cite{Kuperberg}. Our exposition is 
elementary in the hope that it will serve as an introduction 
to this part of exactly solved statistical mechanical models 
for non-physicists.

\section{The model}

To be reasonably self-contained, we start by recalling basic 
definitions related to lattice models in general, and to the 
dwbc six vertex model in particular. We refer the reader to 
\cite{Baxter-book,Korepin-book} for further details. 

\paragraph*{Lattice configurations} Consider a square lattice, 
with $N_c$ vertical lines (columns) labeled {\it 
from left to right} as $\{N_c, N_c - 1, \cdots, 2, 1\}$, and 
$N_r$ horizontal lines (rows) labeled {\it from 
top to bottom} as $\{1, 2, \cdots, N_r - 1, N_r\}$.
Initially, we take $N_c = N_r = N$. Later, we will relax 
this condition and consider lattices with deviations from
dwbc's and $N_c \ne N_r$.


\begin{center}
\begin{minipage}{2in}

\setlength{\unitlength}{0.001cm}

\begin{picture}(5000, 4500)(0, 0)
\thicklines

\path(0600, 3600)(4200, 3600)
\path(0600, 3000)(4200, 3000)
\path(0600, 2400)(4200, 2400)
\path(0600, 1800)(4200, 1800)
\path(0600, 1200)(4200, 1200)

\path(1200, 4200)(1200, 0600)
\path(1800, 4200)(1800, 0600)
\path(2400, 4200)(2400, 0600)
\path(3000, 4200)(3000, 0600)
\path(3600, 4200)(3600, 0600)

\put(0100, 3600){$1$}
\put(0100, 3000){$2$}
\put(0100, 2400){$3$}

\put(0100, 1200){$N$}

\put(1100, 0100){$N$}

\put(2300, 0100){$3$}
\put(2900, 0100){$2$}
\put(3500, 0100){$1$}

\end{picture}

\begin{ca}
\label{1figure}
Our playing field
\end{ca}

\end{minipage}
\end{center}

\paragraph*{Boundary lines} or simply `boundaries', are the 
right most, left most, top and bottom lattice lines.

\paragraph*{Orientations} We assign each lattice line 
an orientation. In our convention, horizontal lines are 
oriented {\it from left to right}. Vertical lines are 
oriented {\it from bottom to top}. 


\begin{center}
\begin{minipage}{4in}

\setlength{\unitlength}{0.001cm}
\begin{picture}(5000, 5000)(-2000, 0)
\thicklines

\path(1200,4200)(4800,4200)
\path(1200,3600)(4800,3600)
\path(1200,3000)(4800,3000)
\path(1200,2400)(4800,2400)
\path(1200,1800)(4800,1800)

\path(1800,4800)(1800,1200)
\path(2400,4800)(2400,1200)
\path(3000,4800)(3000,1200)
\path(3600,4800)(3600,1200)
\path(4200,4800)(4200,1200)

\path(00,3600)(600,3600)
\whiten\path(240,3510)(600,3600)(240,3690)(240,3510)

\path(00,3000)(600,3000)
\whiten\path(240,2910)(600,3000)(240,3090)(240,2910)

\path(00,4200)(600,4200)
\whiten\path(240,4110)(600,4200)(240,4290)(240,4110)

\path(00,2400)(600,2400)
\whiten\path(240,2310)(600,2400)(240,2490)(240,2310)

\path(00,1800)(600,1800)
\whiten\path(240,1710)(600,1800)(240,1890)(240,1710)

\path(1800,00)(1800,600)
\whiten\path(1890,240)(1800,600)(1710,240)(1890,240)

\path(2400,00)(2400,600)
\whiten\path(2490,240)(2400,600)(2310,240)(2490,240)

\path(3000,00)(3000,600)
\whiten\path(3090,240)(3000,600)(2910,240)(3090,240)

\path(3600,00)(3600,600)
\whiten\path(3690,240)(3600,600)(3510,240)(3690,240)

\path(4200,00)(4200,600)
\whiten\path(4290,240)(4200,600)(4110,240)(4290,240)

\end{picture}

\begin{ca}
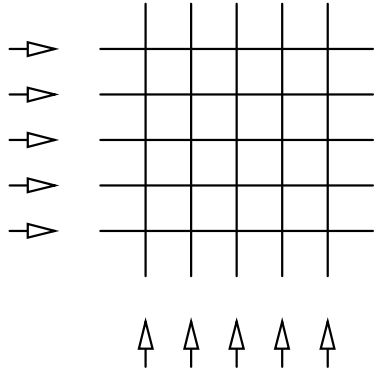
The white arrows denote the orientations of the lattice 
lines.
\end{ca}

\end{minipage}
\end{center}

\paragraph*{Rapidities} We assign each oriented lattice line 
a complex variable called a rapidity. 
We assign the {\it i-th} horizontal line a rapidity $x_i$, 
and the {\it j-th} vertical line a rapidity $y_j$. 
We restrict our attention to the fully inhomogeneous 
situation in which all rapidities are different.  


\begin{center}
\begin{minipage}{4in}

\setlength{\unitlength}{0.001cm}
\begin{picture}(5000, 6000)(-1500, 0)
\thicklines

\path(2400,5400)(2400,1800)
\path(3000,5400)(3000,1800)
\path(3600,5400)(3600,1800)
\path(4200,5400)(4200,1800)
\path(4800,5400)(4800,1800)

\path(1800,4800)(5400,4800)
\path(1800,4200)(5400,4200)
\path(1800,3600)(5400,3600)
\path(1800,3000)(5400,3000)
\path(1800,2400)(5400,2400)

\path(0600,4254)(1200,4254)
\whiten\path(840,4164)(1200,4254)(840,4344)(840,4164)
\path(600,3654)(1200,3654)
\whiten\path(840,3564)(1200,3654)(840,3744)(840,3564)
\path(600,4854)(1200,4854)
\whiten\path(840,4764)(1200,4854)(840,4944)(840,4764)
\path(600,3054)(1200,3054)
\whiten\path(840,2964)(1200,3054)(840,3144)(840,2964)
\path(600,2454)(1200,2454)
\whiten\path(840,2364)(1200,2454)(840,2544)(840,2364)
\path(2400,654)(2400,1254)
\whiten\path(2490,894)(2400,1254)(2310,894)(2490,894)
\path(3000,654)(3000,1254)
\whiten\path(3090,894)(3000,1254)(2910,894)(3090,894)
\path(3600,654)(3600,1254)
\whiten\path(3690,894)(3600,1254)(3510,894)(3690,894)
\path(4200,654)(4200,1254)
\whiten\path(4290,894)(4200,1254)(4110,894)(4290,894)
\path(4800,654)(4800,1254)
\whiten\path(4890,894)(4800,1254)(4710,894)(4890,894)

\put(0100,4854){$x_1$}
\put(0100,4254){$x_2$}

\put(0100,2454){$x_N$}

\put(2300,0250){$y_N$}

\put(4100,0250){$y_2$}
\put(4700,0250){$y_1$}

\end{picture}

\begin{ca}
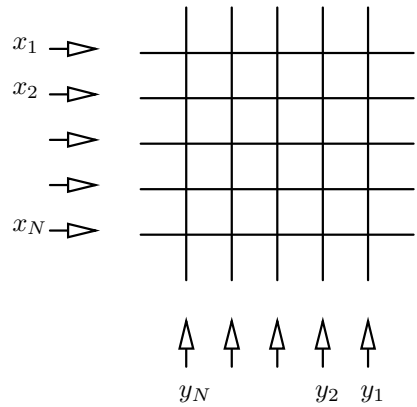
The rapidity flows along the lattice lines. The direction of the 
flow is the orientation of the line. The strength of the flow is 
the rapidity. In our convention, the flow in each line orthogonally 
crosses lines with decreasing label. 
\end{ca}

\end{minipage}
\end{center}

\paragraph*{Bonds} are horizontal or vertical line segments that 
lies between two intersection points, or at the very end of a 
line.

\paragraph*{Arrows} To each bond we assign an arrow that can point 
in either direction along the bond. 

\paragraph*{Vertices} We call the intersection point of the
$i$-th horizontal line and the $j$-th vertical line, together 
with the 4 bonds attached to it, and the arrows on these bonds, 
a vertex $v_{ij}$. 

This extends to the general case when the two intersecting lines 
are not `everywhere' straight. The line orientations are sufficient 
to determine which of the two intersecting lines is `locally' 
vertical and the other is `locally' horizontal\footnote{When in 
doubt, deform the intersecting lines locally such that their 
orientations agree with those given in figure \ref{sixverticesfigure}.}.

\paragraph*{Weights} We assign each vertex, $v_{ij}$, a (Boltzmann) 
weight, $W_{ij}$, that depends on the following data 

\begin{enumerate}
\item The orientations of the lines that intersect at that vertex.
      If necessary, we must rotate a vertex so that one line is 
      oriented from left to right, while the other is oriented
      from bottom to top.
      
\item The orientations of the arrows on the four bonds attached
      to the intersection point. 

\item The rapidities, $\{x_i, y_j\}$, flowing through the 
      intersection point. 

\item A parameter $\lambda$, called the crossing parameter that 
      is the same for all vertices $v_{ij}$, and characterises
      the physical characteristics of the model. 
\end{enumerate}

\paragraph*{The six vertex model} Given that each vertex has 4
bonds, and each bond carries an arrow with 2 possible orientations, 
there are 16 possible vertex configurations. 
The six vertex model is the special case where only six types of
vertices generally have non-zero weights. The weights of the 
remaining 10 types are set to zero. 
weights.


\begin{center}
\begin{minipage}{4.3in}

\setlength{\unitlength}{0.0008cm}
\begin{picture}(10000,7500)(-1000, 0)
\thicklines

\blacken\path(10162,2310)(10522,2400)(10162,2490)(10162,2310)
\blacken\path(1282,2490)(0922,2400)(1282,2310)(1282,2490)
\blacken\path(1462,5610)(1822,5700)(1462,5790)(1462,5610)
\blacken\path(1732,1860)(1822,1500)(1912,1860)(1732,1860)
\blacken\path(1732,2685)(1822,2325)(1912,2685)(1732,2685)
\blacken\path(1912,5340)(1822,5700)(1732,5340)(1912,5340)
\blacken\path(1912,6240)(1822,6600)(1732,6240)(1912,6240)
\blacken\path(2182,2490)(1822,2400)(2182,2310)(2182,2490)
\blacken\path(2362,5610)(2722,5700)(2362,5790)(2362,5610)
\blacken\path(5182,5790)(4822,5700)(5182,5610)(5182,5790)
\blacken\path(5362,2310)(5722,2400)(5362,2490)(5362,2310)
\blacken\path(5632,1860)(5722,1500)(5812,1860)(5632,1860)
\blacken\path(5632,2760)(5722,2400)(5812,2760)(5632,2760)
\blacken\path(5812,5340)(5722,5700)(5632,5340)(5812,5340)
\blacken\path(5812,6240)(5722,6600)(5632,6240)(5812,6240)
\blacken\path(6082,5790)(5722,5700)(6082,5610)(6082,5790)
\blacken\path(6262,2310)(6622,2400)(6262,2490)(6262,2310)
\blacken\path(9082,2490)(8722,2400)(9082,2310)(9082,2490)
\blacken\path(9262,5610)(9622,5700)(9262,5790)(9262,5610)
\blacken\path(9532,2760)(9622,2400)(9712,2760)(9532,2760)
\blacken\path(9532,5160)(9622,4800)(9712,5160)(9532,5160)
\blacken\path(9712,2040)(9622,2400)(9532,2040)(9712,2040)
\blacken\path(9712,6240)(9622,6600)(9532,6240)(9712,6240)
\blacken\path(9982,5790)(9622,5700)(9982,5610)(9982,5790)
\path(10222,5700)(9622,5700)
\path(1222,5700)(1822,5700)
\path(1522,2400)(0922,2400)
\path(1822,1500)(1822,2100)
\path(1822,2325)(1822,2925)
\path(1822,3300)(1822,1500)
\path(1822,3900)(1822,4500)
\path(1822,5700)(1822,5100)
\path(1822,0600)(1822,1200)
\path(1822,6600)(1822,4800)
\path(0022,2400)(0622,2400)
\path(0022,5700)(0622,5700)
\path(2422,2400)(1822,2400)
\path(3922,2400)(4522,2400)
\path(3922,5700)(4522,5700)
\path(4822,2400)(6622,2400)
\path(4822,5700)(6622,5700)
\path(5122,2400)(5722,2400)
\path(5422,5700)(4822,5700)
\path(5722,1500)(5722,2100)
\path(5722,2400)(5722,3000)
\path(5722,3300)(5722,1500)
\path(5722,3900)(5722,4500)
\path(5722,5700)(5722,5100)
\path(5722,600)(5722,1200)
\path(5722,6600)(5722,4800)
\path(5722,6600)(5722,6000)
\path(6022,2400)(6622,2400)
\path(6322,5700)(5722,5700)
\path(7822,2400)(8422,2400)
\path(7822,5700)(8422,5700)
\path(8722,2400)(10522,2400)
\path(8722,5700)(10522,5700)
\path(9022,5700)(9622,5700)
\path(0922,2400)(2722,2400)
\path(0922,5700)(2722,5700)
\path(9322,2400)(8722,2400)
\path(9622,2400)(9622,1800)
\path(9622,2400)(9622,3000)
\path(9622,3300)(9622,1500)
\path(9622,3900)(9622,4500)
\path(9622,4800)(9622,5400)
\path(9622,0600)(9622,1200)
\path(9622,6600)(9622,4800)
\path(9622,6600)(9622,6000)
\path(9922,2400)(10522,2400)

\whiten\path(1912,4140)(1822,4500)(1732,4140)(1912,4140)
\whiten\path(1912,0840)(1822,1200)(1732,0840)(1912,0840)
\whiten\path(0262,2310)(0622,2400)(0262,2490)(0262,2310)
\whiten\path(0262,5610)(0622,5700)(0262,5790)(0262,5610)
\whiten\path(4162,2310)(4522,2400)(4162,2490)(4162,2310)
\whiten\path(4162,5610)(4522,5700)(4162,5790)(4162,5610)
\whiten\path(5812,4140)(5722,4500)(5632,4140)(5812,4140)
\whiten\path(5812,0840)(5722,1200)(5632,0840)(5812,0840)
\whiten\path(8062,2310)(8422,2400)(8062,2490)(8062,2310)
\whiten\path(8062,5610)(8422,5700)(8062,5790)(8062,5610)
\whiten\path(9712,4140)(9622,4500)(9532,4140)(9712,4140)
\whiten\path(9712,0840)(9622,1200)(9532,0840)(9712,0840)

\put(1000,0000){$a(x, y)$}
\put(5000,0000){$b(x, y)$}
\put(9000,0000){$c(x, y)$}

\put(0300,5000){$x$}
\put(4200,5000){$x$}
\put(8100,5000){$x$}

\put(0300,1700){$x$}
\put(4200,1700){$x$}
\put(8100,1700){$x$}

\put(2000,4200){$y$}
\put(6000,4200){$y$}
\put(10000,4200){$y$}

\put(2000,0900){$y$}
\put(6000,0900){$y$}
\put(10000,0900){$y$}

\end{picture}

\begin{ca}
\label{sixverticesfigure}
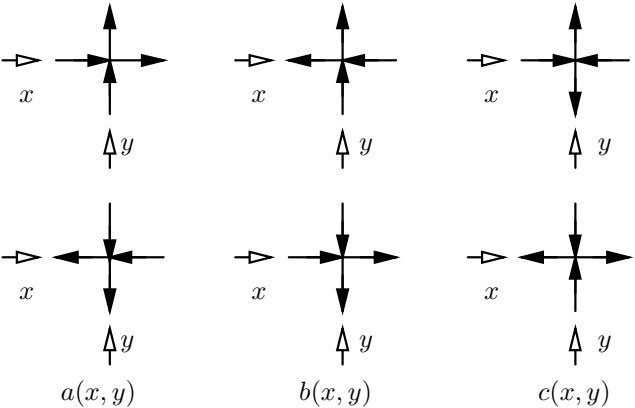
The vertices of the six vertex model, with their rapidity flows.
The weights of every two vertices in the same column are equal,
and is shown below it.
\end{ca}

\end{minipage}
\end{center}

\paragraph*{Conservation of arrow flow} The vertices with non-zero 
weights are precisely those that conserve arrow flow: In each vertex, 
two arrows point inwards, and two points outwards. 

\paragraph*{Bracket notation} We define $[x] = \sinh(\lambda x)$, 
where $\lambda$ is the crossing parameter. 

\paragraph*{Vertex weights} In the bracket notation, the vertex 
weights are

\begin{equation}
\label{weightsequation}
a(x_i, y_j) = [-x_i + y_j + 1],
\quad
b(x_i, y_j) = [-x_i + y_j    ],
\quad
c(x_i, y_j) = [1]
\end{equation}
where $x_i$ (respectively $y_i$) is the horizontal (respectively 
vertical) rapidity variable flowing through the vertex. 

The weights of type $a$ and type $b$ vertices depend on the 
differences of the rapidities flowing through the vertex, 
and can vanish for specific values of the rapidities. The 
weight of the $c$ vertex is independent of the rapidities.

\paragraph*{Yang Baxter equations} The origin of solvability 
of the six vertex model (irrespective of the boundary conditions) 
is that the (Boltzmann) weights of the model satisfy the Yang Baxter 
equations \cite{Baxter-book}. Defining the ($R$) matrix

\begin{equation}
\label{rmatrixequation}
R(x, y) =
\ll
\begin{array}{cccc}
a(x, y) & 0       & 0        & 0        \\
0       & b(x, y) & c(x, y)  & 0        \\
0       & c(x, y) & b(x, y)  & 0        \\
0       & 0       & 0        & a(x, y)
\end{array}
\rr
\end{equation}
the set of all Yang Baxter equations can be written in matrix 
form as 

\begin{equation}
\label{RRRequation}
R(x, y) 
R(x, z) 
R(y, z) 
=
R(y, z)
R(x, z) 
R(x, y)
\end{equation}
where matrix multiplication is implied. The matrix equation 
\ref{RRRequation} is equivalent to a set of equations between 
the weights. An example of a Yang Baxter equation, that will
be used below, is shown graphically in figure \ref{yangbaxterfigure}. 


\begin{center}
\begin{minipage}{4.5in}

\setlength{\unitlength}{0.0008cm}

\begin{picture}(10000,6000)(-3000, 0)

\thicklines

\blacken\path(202,4762)(112,5122)(22,4762)(202,4762)
\path(112,5122)(112,3622)
\path(1912,3622)(1912,2722)
\path(1012,1822)(1012,1222)
\blacken\path(922,1582)(1012,1222)(1102,1582)(922,1582)
\path(4912,3622)(4912,2722)
\path(6112,3622)(6112,2722)
\blacken\path(3202,4762)(3112,5122)(3022,4762)(3202,4762)
\path(3112,5122)(3112,3622)
\blacken\path(4102,4762)(4012,5122)(3922,4762)(4102,4762)
\path(4012,5122)(4012,4522)
\blacken\path(6202,4762)(6112,5122)(6022,4762)(6202,4762)
\path(6112,5122)(6112,4522)
\blacken\path(7102,4762)(7012,5122)(6922,4762)(7102,4762)
\path(7012,5122)(7012,4522)
\path(7012,1822)(7012,1222)
\blacken\path(6922,1582)(7012,1222)(7102,1582)(6922,1582)
\path(4012,1822)(4012,1222)
\blacken\path(3922,1582)(4012,1222)(4102,1582)(3922,1582)
\blacken\path(1822,3382)(1912,3022)(2002,3382)(1822,3382)
\path(1912,3022)(1912,3322)
\blacken\path(6202,2962)(6112,3322)(6022,2962)(6202,2962)
\path(6112,3322)(6112,3022)
\blacken\path(1102,4762)(1012,5122)(922,4762)(1102,4762)
\path(1012,5122)(1012,4522)
\path(1912,4522)(562,3172)
\path(1012,4522)(1912,3622)
\path(112,3622)(562,3172)
\path(562,3172)(112,2722)
\path(562,3172)(1462,2272)
\path(1912,2722)(1012,1822)
\path(3112,3622)(3562,3172)
\path(4462,4072)(3562,3172)
\path(3562,3172)(3112,2722)
\path(3562,3172)(4462,2272)
\path(4912,2722)(4012,1822)
\path(6112,2722)(7012,1822)
\path(7462,3172)(6112,1822)
\path(6562,4072)(7912,2722)
\path(7912,3622)(7462,3172)
\path(7012,4522)(6112,3622)
\path(4912,4522)(4462,4072)
\path(4012,4522)(4462,4072)
\path(6112,4522)(6562,4072)
\blacken\path(5002,2962)(4912,3322)(4822,2962)(5002,2962)
\path(4912,3322)(4912,3022)
\path(4462,4072)(4912,3622)
\blacken\path(7067.198,3694.081)(6749,3885)(6939.919,3566.802)(7067.198,3694.081)
\path(6749,3885)(7199,3435)
\blacken\path(3977.919,3715.198)(3787,3397)(4105.198,3587.919)(3977.919,3715.198)
\path(3787,3397)(4237,3847)
\blacken\path(3955.802,2650.919)(4274,2460)(4083.081,2778.198)(3955.802,2650.919)
\path(4274,2460)(3824,2910)
\blacken\path(1046.081,3528.802)(1237,3847)(918.802,3656.081)(1046.081,3528.802)
\path(1237,3847)(787,3397)
\blacken\path(1105.198,2756.081)(787,2947)(977.919,2628.802)(1105.198,2756.081)
\path(787,2947)(1237,2497)
\blacken\path(202,1312)(112,1672)(22,1312)(202,1312)
\path(112,1672)(112,1372)
\path(112,2722)(112,1222)
\path(1462,2272)(1912,1822)
\path(4462,2272)(4912,1822)
\path(4912,1822)(4912,1222)
\blacken\path(5002,1312)(4912,1672)(4822,1312)(5002,1312)
\path(4912,1672)(4912,1372)
\blacken\path(2002,1312)(1912,1672)(1822,1312)(2002,1312)
\path(1912,1672)(1912,1372)
\blacken\path(3202,1312)(3112,1672)(3022,1312)(3202,1312)
\path(3112,1672)(3112,1372)
\path(3112,2722)(3112,1222)
\path(1912,1822)(1912,1297)(1912,1222)
\blacken\path(8002,1312)(7912,1672)(7822,1312)(8002,1312)
\path(7912,1672)(7912,1372)
\path(7912,2722)(7912,1222)
\path(6112,1822)(6112,1222)
\blacken\path(6202,1312)(6112,1672)(6022,1312)(6202,1312)
\path(6112,1672)(6112,1372)
\blacken\path(7822,4982)(7912,4622)(8002,4982)(7822,4982)
\path(7912,4622)(7912,4922)
\blacken\path(4822,4982)(4912,4622)(5002,4982)(4822,4982)
\path(4912,4622)(4912,4922)
\blacken\path(1822,4982)(1912,4622)(2002,4982)(1822,4982)
\path(1912,4622)(1912,4922)
\path(7912,5122)(7912,3622)
\path(4912,5122)(4912,4522)
\path(1912,5122)(1912,4522)
\blacken\path(6977.919,2815.198)(6787,2497)(7105.198,2687.919)(6977.919,2815.198)
\path(6787,2497)(7237,2947)
\path(112,22)(112,622)
\whiten\path(202,262)(112,622)(22,262)(202,262)
\path(1012,22)(1012,622)
\whiten\path(1102,262)(1012,622)(922,262)(1102,262)
\path(1912,22)(1912,622)
\whiten\path(2002,262)(1912,622)(1822,262)(2002,262)
\path(3112,22)(3112,622)
\whiten\path(3202,262)(3112,622)(3022,262)(3202,262)
\path(4012,22)(4012,622)
\whiten\path(4102,262)(4012,622)(3922,262)(4102,262)
\path(4912,22)(4912,622)
\whiten\path(5002,262)(4912,622)(4822,262)(5002,262)
\path(6112,22)(6112,622)
\whiten\path(6202,262)(6112,622)(6022,262)(6202,262)
\path(7012,22)(7012,622)
\whiten\path(7102,262)(7012,622)(6922,262)(7102,262)
\path(7912,22)(7912,622)
\whiten\path(8002,262)(7912,622)(7822,262)(8002,262)

\put(2400,3000){$+$}
\put(5400,3000){$=$}
\put(0000,0900){$x$}
\put(0900,0900){$y$}
\put(1800,0900){$z$}
\put(3000,0900){$x$}
\put(3900,0900){$y$}
\put(4800,0900){$z$}
\put(6000,0900){$x$}
\put(6900,0900){$y$}
\put(7800,0900){$z$}
\put(0000,5400){$z$}
\put(0900,5400){$y$}
\put(1800,5400){$x$}
\put(3000,5400){$z$}
\put(3900,5400){$y$}
\put(4800,5400){$x$}
\put(6000,5400){$z$}
\put(6900,5400){$y$}
\put(7800,5400){$x$}

\end{picture}

\begin{ca}
\label{yangbaxterfigure}
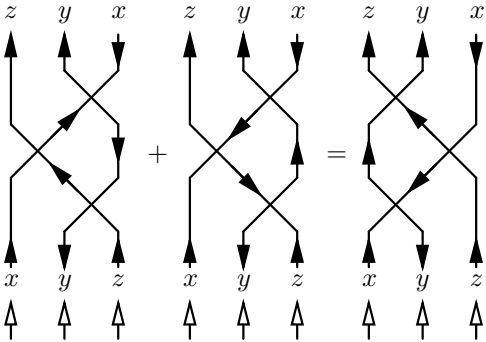
A Yang Baxter equation. The external arrows, rapidities, and 
line orientations are the same on corresponding lines.  
Line orientations determine the vertex type unambiguously. 
\end{ca}

\end{minipage}
\end{center}

The Yang Baxter equation corresponding to figure 
\ref{yangbaxterfigure} is

\begin{equation}
\label{yangbaxterequation}
b(y, z) a(x, z) c(x, y)
+
c(y, z) c(x, z) b(x, y)
=
c(x, y) b(x, z) a(y, z)
\end{equation}

\paragraph*{Domain wall boundary conditions (dwbc's)} For 
a vertex model on a finite square lattice, with $N_c = N_r = N$, 
one can require 
that all arrows on the left and right boundaries point inwards,  
and all arrows on the upper and lower boundaries point outwards. 
These are called `domain wall boundary conditions' 
(dwbc's)\footnote{One can equally well make the opposite choice 
of boundary arrow orientations. For consistency, we need to be 
make one choice and stay with that. This work is full of such
choices.}. An example is shown in figure \ref{dwbcfigure}. 


\begin{center}
\begin{minipage}{4in}

\setlength{\unitlength}{0.001cm}

\begin{picture}(6000,6000)(-1000, 0)
\thicklines

\path(300,4800)(5700,4800)
\path(300,3900)(5700,3900)
\path(300,2100)(5700,2100)
\path(300,1200)(5700,1200)
\path(1200,5700)(1200,300)
\path(2100,5700)(2100,300)
\path(4800,5700)(4800,300)
\path(1200,5025)(1200,5625)
\blacken\path(1290,5265)(1200,5625)(1110,5265)(1290,5265)
\path(2100,5025)(2100,5625)
\blacken\path(2190,5265)(2100,5625)(2010,5265)(2190,5265)
\path(3900,5025)(3900,5625)
\blacken\path(3990,5265)(3900,5625)(3810,5265)(3990,5265)
\path(4800,5100)(4800,5700)
\blacken\path(4905,5340)(4800,5700)(4695,5340)(4905,5340)
\path(1425,4800)(2025,4800)
\blacken\path(1665,4710)(2025,4800)(1665,4890)(1665,4710)
\path(3675,4800)(3075,4800)
\blacken\path(3435,4890)(3075,4800)(3435,4710)(3435,4890)
\path(4500,4800)(3900,4800)
\blacken\path(4260,4890)(3900,4800)(4260,4710)(4260,4890)
\path(5400,4800)(4800,4800)
\blacken\path(5160,4890)(4800,4800)(5160,4710)(5160,4890)
\path(525,3900)(1125,3900)
\blacken\path(765,3810)(1125,3900)(765,3990)(765,3810)
\path(1800,3900)(1200,3900)
\blacken\path(1560,3990)(1200,3900)(1560,3810)(1560,3990)
\path(5400,3900)(4800,3900)
\blacken\path(5160,3990)(4800,3900)(5160,3810)(5160,3990)
\path(600,3000)(1200,3000)
\blacken\path(840,2910)(1200,3000)(840,3090)(840,2910)
\path(600,2100)(1200,2100)
\blacken\path(840,2010)(1200,2100)(840,2190)(840,2010)
\path(600,1200)(1200,1200)
\blacken\path(840,1110)(1200,1200)(840,1290)(840,1110)
\path(1425,3000)(2025,3000)
\blacken\path(1665,2910)(2025,3000)(1665,3090)(1665,2910)
\path(3600,3000)(3000,3000)
\blacken\path(3360,3090)(3000,3000)(3360,2910)(3360,3090)
\path(4575,3000)(3975,3000)
\blacken\path(4335,3090)(3975,3000)(4335,2910)(4335,3090)
\path(5475,3000)(4875,3000)
\blacken\path(5235,3090)(4875,3000)(5235,2910)(5235,3090)
\path(5400,2100)(4800,2100)
\blacken\path(5160,2190)(4800,2100)(5160,2010)(5160,2190)
\path(5475,1200)(4875,1200)
\blacken\path(5235,1290)(4875,1200)(5235,1110)(5235,1290)
\path(4800,1800)(4800,1200)
\blacken\path(4710,1560)(4800,1200)(4890,1560)(4710,1560)
\path(1500,1200)(2100,1200)
\blacken\path(1740,1110)(2100,1200)(1740,1290)(1740,1110)
\path(4575,1200)(3975,1200)
\blacken\path(4335,1290)(3975,1200)(4335,1110)(4335,1290)
\path(1200,975)(1200,375)
\blacken\path(1110,735)(1200,375)(1290,735)(1110,735)
\path(2100,900)(2100,300)
\blacken\path(2010,660)(2100,300)(2190,660)(2010,660)
\path(3900,900)(3900,300)
\blacken\path(3810,660)(3900,300)(3990,660)(3810,660)
\path(4800,900)(4800,300)
\blacken\path(4710,660)(4800,300)(4890,660)(4710,660)
\path(525,4800)(1125,4800)
\blacken\path(765,4710)(1125,4800)(765,4890)(765,4710)
\path(4200,2100)(4800,2100)
\blacken\path(4440,2010)(4800,2100)(4440,2190)(4440,2010)
\path(4800,2400)(4800,3000)
\blacken\path(4890,2640)(4800,3000)(4710,2640)(4890,2640)
\path(4800,3375)(4800,3975)
\blacken\path(4890,3615)(4800,3975)(4710,3615)(4890,3615)
\path(4800,4200)(4800,4800)
\blacken\path(4890,4440)(4800,4800)(4710,4440)(4890,4440)
\path(4500,3900)(3900,3900)
\blacken\path(4260,3990)(3900,3900)(4260,3810)(4260,3990)
\path(3000,5700)(3000,300)
\path(3900,4200)(3900,4800)
\blacken\path(3990,4440)(3900,4800)(3810,4440)(3990,4440)
\path(3900,5700)(3900,300)
\path(3300,3900)(3900,3900)
\blacken\path(3540,3810)(3900,3900)(3540,3990)(3540,3810)
\path(3900,3600)(3900,3000)
\blacken\path(3810,3360)(3900,3000)(3990,3360)(3810,3360)
\path(3900,2700)(3900,2100)
\blacken\path(3810,2460)(3900,2100)(3990,2460)(3810,2460)
\path(3000,5100)(3000,5700)
\blacken\path(3090,5340)(3000,5700)(2910,5340)(3090,5340)
\path(3000,4500)(3000,3900)
\blacken\path(2910,4260)(3000,3900)(3090,4260)(2910,4260)
\path(3000,3300)(3000,3900)
\blacken\path(3090,3540)(3000,3900)(2910,3540)(3090,3540)
\path(3000,2400)(3000,3000)
\blacken\path(3090,2640)(3000,3000)(2910,2640)(3090,2640)
\path(3000,900)(3000,300)
\blacken\path(2910,660)(3000,300)(3090,660)(2910,660)
\path(2100,2700)(2100,2100)
\blacken\path(2010,2460)(2100,2100)(2190,2460)(2010,2460)
\path(2100,3375)(2100,3975)
\blacken\path(2190,3615)(2100,3975)(2010,3615)(2190,3615)
\path(2100,4200)(2100,4800)
\blacken\path(2190,4440)(2100,4800)(2010,4440)(2190,4440)
\path(1200,4275)(1200,4875)
\blacken\path(1290,4515)(1200,4875)(1110,4515)(1290,4515)
\path(1200,3600)(1200,3000)
\blacken\path(1110,3360)(1200,3000)(1290,3360)(1110,3360)
\path(1200,2700)(1200,2100)
\blacken\path(1110,2460)(1200,2100)(1290,2460)(1110,2460)
\path(1200,1800)(1200,1200)
\blacken\path(1110,1560)(1200,1200)(1290,1560)(1110,1560)
\path(3300,2100)(3900,2100)
\blacken\path(3540,2010)(3900,2100)(3540,2190)(3540,2010)
\path(3900,1800)(3900,1200)
\blacken\path(3810,1560)(3900,1200)(3990,1560)(3810,1560)
\path(3600,1200)(3000,1200)
\blacken\path(3360,1290)(3000,1200)(3360,1110)(3360,1290)
\path(300,3000)(5700,3000)
\path(2400,4800)(3000,4800)
\blacken\path(2640,4710)(3000,4800)(2640,4890)(2640,4710)
\path(2625,3900)(2025,3900)
\blacken\path(2385,3990)(2025,3900)(2385,3810)(2385,3990)
\path(2700,3000)(2100,3000)
\blacken\path(2460,3090)(2100,3000)(2460,2910)(2460,3090)
\path(3000,1500)(3000,2100)
\blacken\path(3090,1740)(3000,2100)(2910,1740)(3090,1740)
\path(2400,2100)(3000,2100)
\blacken\path(2640,2010)(3000,2100)(2640,2190)(2640,2010)
\path(2400,1200)(3000,1200)
\blacken\path(2640,1110)(3000,1200)(2640,1290)(2640,1110)
\path(1500,2100)(2100,2100)
\blacken\path(1740,2010)(2100,2100)(1740,2190)(1740,2010)
\path(2100,1800)(2100,1200)
\blacken\path(2010,1560)(2100,1200)(2190,1560)(2010,1560)

\put(0,4800){$1$}
\put(0,3900){$2$}
\put(0,3000){$3$}
\put(0,1200){$N$}

\put(4700,-200){$1$}
\put(3800,-200){$2$}
\put(2900,-200){$3$}
\put(1100,-200){$N$}
\end{picture}

\begin{ca}
\label{dwbcfigure}
A domain wall boundary configuration.
\end{ca}
\end{minipage}
\end{center}

\paragraph*{Partition function} Given a statistical mechanical
model, the partition function is defined as the sum over 
{\it all} weighted configurations. The weight of each
configuration is the product of the weights, 
$W_{ij}$, of its vertices $v_{ij}$

\begin{equation}
\label{partitionfunctionequation}
Z_{N}\ll \{x\}, \{y\}\rr = 
\sum_{{\rm all}{\phantom{-}}{\rm allowed} 
       \atop  
      {\rm configurations}}
\prod_{\rm vertices} W_{ij}
\end{equation}

\begin{re}
The expression in equation \ref{partitionfunctionequation} is 
computationally worthless, as it cannot be evaluated in polynomial 
time in $N$.
\end{re}

\subsection{Izergin's expression for the partition function}

In \cite{Izergin}, Izergin solved Korepin's recursion relations
\cite{Korepin}, and obtained an explicit expression for the 
partition function of the dwbc six vertex model in terms of 
a determinant

$$
Z_N\ll \{x\}, \{y\}\rr 
=
\frac{\prod_{{i, j}=1}^{N}      a(x_i, y_j) b(x_i, y_j)}
     {\prod_{1 \le i < j \le N} b(x_i, x_j) b(y_j, y_i)}
     \, det M 
$$
\begin{equation}
\label{izerginequation}
\phantom{Z_N\ll \{x\}, \{y\}\rr}
=
\frac{\prod_{{i, j}=1}^{N} [- x_i + y_j + 1][- x_i + y_j]}
     {\prod_{1 \le i < j \le N}[- x_i + x_j][- y_j + y_i]} 
     \, det M 
\end{equation}
where
$$
M_{ij} = \frac{c(x_i, y_j)}{a(x_i, y_j) b(x_i, y_j)}
       = \frac{[1]}{[- x_{i}+ y_{j}+1][-x_{i}+y_{j}]}
$$

\begin{re}
Izergin's expression, equation \ref{izerginequation} for the 
partition function in terms of a determinant is computationally 
meaningful, as it can be evaluated in polynomial time in $N$.
\end{re}

\section{Correlation functions} 

Apart from the partition function which is the weighted sum 
over {\it all} configurations allowed by the dwbc's, one wishes 
to consider $n$-point correlation functions. 
These are defined as weighted sums over all configurations 
such that $n$ arrows (that are summed over in the partition 
function) have certain frozen orientations. 

Since the arrows on the outside of all vertical and 
horizontal lines are always frozen by definition of 
dwbc's, we need to consider 
freezing arrows that are on the inside of the lattice. 

We start by considering the 1-point function obtained 
by summing all configurations such that a certain single 
arrow is kept frozen. Without loss of generality, let us 
consider freezing a horizontal arrow. 

For example, consider freezing the orientation of the white 
arrow in figure \ref{bulkfigure}, where bonds without arrows
indicate bonds whose arrow orientations are summed over


\begin{center}
\begin{minipage}{2.5in}
\setlength{\unitlength}{0.001cm}
\begin{picture}(6000,6000)(500, 0)

\thicklines

\path(300,4800)(5700,4800)
\path(300,3900)(5700,3900)
\path(300,1200)(5700,1200)
\path(2100,5700)(2100,300)
\path(4800,5700)(4800,300)
\path(2100,5025)(2100,5625)
\blacken\path(2190,5265)(2100,5625)(2010,5265)(2190,5265)
\path(4800,5100)(4800,5700)
\blacken\path(4905,5340)(4800,5700)(4695,5340)(4905,5340)
\path(5400,4800)(4800,4800)
\blacken\path(5160,4890)(4800,4800)(5160,4710)(5160,4890)
\path(525,3900)(1125,3900)
\blacken\path(765,3810)(1125,3900)(765,3990)(765,3810)
\path(5400,3900)(4800,3900)
\blacken\path(5160,3990)(4800,3900)(5160,3810)(5160,3990)
\path(600,3000)(1200,3000)
\blacken\path(840,2910)(1200,3000)(840,3090)(840,2910)
\path(600,2100)(1200,2100)
\blacken\path(840,2010)(1200,2100)(840,2190)(840,2010)
\path(600,1200)(1200,1200)
\blacken\path(840,1110)(1200,1200)(840,1290)(840,1110)
\path(5475,3000)(4875,3000)
\blacken\path(5235,3090)(4875,3000)(5235,2910)(5235,3090)
\path(5400,2100)(4800,2100)
\blacken\path(5160,2190)(4800,2100)(5160,2010)(5160,2190)
\path(5475,1200)(4875,1200)
\blacken\path(5235,1290)(4875,1200)(5235,1110)(5235,1290)
\path(2100,900)(2100,300)
\blacken\path(2010,660)(2100,300)(2190,660)(2010,660)
\path(3900,900)(3900,300)
\blacken\path(3810,660)(3900,300)(3990,660)(3810,660)
\path(4800,900)(4800,300)
\blacken\path(4710,660)(4800,300)(4890,660)(4710,660)
\path(525,4800)(1125,4800)
\blacken\path(765,4710)(1125,4800)(765,4890)(765,4710)
\path(3000,5700)(3000,300)
\path(3000,5100)(3000,5700)
\blacken\path(3090,5340)(3000,5700)(2910,5340)(3090,5340)
\path(3000,900)(3000,300)
\blacken\path(2910,660)(3000,300)(3090,660)(2910,660)
\path(3900,5100)(3900,5700)
\blacken\path(3990,5340)(3900,5700)(3810,5340)(3990,5340)
\path(3900,5700)(3900,300)
\path(300,2100)(5700,2100)
\path(300,3000)(5700,3000)
\path(1200,5100)(1200,5700)
\blacken\path(1290,5340)(1200,5700)(1110,5340)(1290,5340)
\path(1200,900)(1200,300)
\blacken\path(1110,660)(1200,300)(1290,660)(1110,660)
\path(1200,5700)(1200,300)
\path(3300,3900)(3900,3900)
\whiten\path(3540,3810)(3900,3900)(3540,3990)(3540,3810)

\put(-100,4700){$1$}
\put(-100,3800){$2$}
\put(-100,2900){$3$}
\put(-100,1100){$N$}

\put(4700,-200){$1$}
\put(3800,-200){$2$}
\put(2900,-200){$3$}
\put(1100,-200){$N$}
\end{picture}

\begin{ca}
\label{bulkfigure}
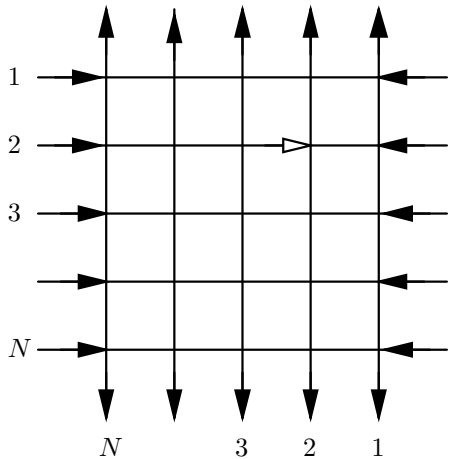
The configurations corresponding to a 1-point function.
\end{ca}

\end{minipage}
\end{center}

Our proposal is that one should start by slicing all such 
configuration into two parts, a right part, and a left 
part


\begin{center}
\begin{minipage}{4in}
\setlength{\unitlength}{0.001cm}

\begin{picture}(7000,6000)(-1000, 0)

\thicklines

\path(2100,5700)(2100,300)
\path(2100,5025)(2100,5625)
\blacken\path(2190,5265)(2100,5625)(2010,5265)(2190,5265)
\path(2100,900)(2100,300)
\blacken\path(2010,660)(2100,300)(2190,660)(2010,660)
\path(3000,5700)(3000,300)
\path(3000,5100)(3000,5700)
\blacken\path(3090,5340)(3000,5700)(2910,5340)(3090,5340)
\path(3000,900)(3000,300)
\blacken\path(2910,660)(3000,300)(3090,660)(2910,660)
\path(1200,5100)(1200,5700)
\blacken\path(1290,5340)(1200,5700)(1110,5340)(1290,5340)
\path(1200,900)(1200,300)
\blacken\path(1110,660)(1200,300)(1290,660)(1110,660)
\path(1200,5700)(1200,300)
\path(300,2100)(3600,2100)
\path(300,1200)(3600,1200)
\path(300,3000)(3600,3000)
\path(300,3900)(3600,3900)
\path(300,4800)(3600,4800)
\path(4200,4800)(6600,4800)
\path(4200,3900)(6600,3900)
\path(4200,3000)(6600,3000)
\path(4200,2100)(6600,2100)
\path(4200,1200)(6600,1200)
\path(4800,5700)(4800,300)
\path(5700,5700)(5700,300)
\path(6300,4800)(5700,4800)
\blacken\path(6060,4890)(5700,4800)(6060,4710)(6060,4890)
\path(6300,3900)(5700,3900)
\blacken\path(6060,3990)(5700,3900)(6060,3810)(6060,3990)
\path(6300,3000)(5700,3000)
\blacken\path(6060,3090)(5700,3000)(6060,2910)(6060,3090)
\path(6300,2100)(5700,2100)
\blacken\path(6060,2190)(5700,2100)(6060,2010)(6060,2190)
\path(6300,1200)(5700,1200)
\blacken\path(6060,1290)(5700,1200)(6060,1110)(6060,1290)
\path(600,4800)(1200,4800)
\blacken\path(840,4710)(1200,4800)(840,4890)(840,4710)
\path(600,3900)(1200,3900)
\blacken\path(840,3810)(1200,3900)(840,3990)(840,3810)
\path(600,3000)(1200,3000)
\blacken\path(840,2910)(1200,3000)(840,3090)(840,2910)
\path(600,2100)(1200,2100)
\blacken\path(840,2010)(1200,2100)(840,2190)(840,2010)
\path(600,1200)(1200,1200)
\blacken\path(840,1110)(1200,1200)(840,1290)(840,1110)
\path(4800,5100)(4800,5700)
\blacken\path(4905,5340)(4800,5700)(4695,5340)(4905,5340)
\path(5700,5100)(5700,5700)
\blacken\path(5805,5340)(5700,5700)(5595,5340)(5805,5340)
\path(4800,900)(4800,300)
\blacken\path(4710,660)(4800,300)(4890,660)(4710,660)
\path(5700,900)(5700,300)
\blacken\path(5610,660)(5700,300)(5790,660)(5610,660)
\thinlines
\path(3150,4950)(3000,4800)
\path(3300,4950)(3000,4650)
\path(3450,4950)(3150,4650)
\path(3600,4950)(3300,4650)
\path(3600,4800)(3450,4650)
\path(3150,3150)(3000,3000)
\path(3300,3150)(3000,2850)
\path(3450,3150)(3150,2850)
\path(3600,3150)(3300,2850)
\path(3600,3000)(3450,2850)
\path(3150,1350)(3000,1200)
\path(3300,1350)(3000,1050)
\path(3450,1350)(3150,1050)
\path(3600,1350)(3300,1050)
\path(3600,1200)(3450,1050)
\path(4350,1350)(4200,1200)
\path(4500,1350)(4200,1050)
\path(4650,1350)(4350,1050)
\path(4800,1350)(4500,1050)
\path(4800,1200)(4650,1050)
\path(4350,3150)(4200,3000)
\path(4500,3150)(4200,2850)
\path(4650,3150)(4350,2850)
\path(4800,3150)(4500,2850)
\path(4800,3000)(4650,2850)
\path(4350,4950)(4200,4800)
\path(4500,4950)(4200,4650)
\path(4650,4950)(4350,4650)
\path(4800,4950)(4500,4650)
\path(4800,4800)(4650,4650)
\path(3150,2250)(3000,2100)
\path(3300,2250)(3000,1950)
\path(3450,2250)(3150,1950)
\path(3600,2250)(3300,1950)
\path(3600,2100)(3450,1950)
\path(4350,2250)(4200,2100)
\path(4500,2250)(4200,1950)
\path(4650,2250)(4350,1950)
\path(4800,2250)(4500,1950)
\path(4800,2100)(4650,1950)
\thicklines
\path(3000,3900)(3600,3900)
\whiten\path(3240,3810)(3600,3900)(3240,3990)(3240,3810)
\path(4200,3900)(4800,3900)
\whiten\path(4440,3810)(4800,3900)(4440,3990)(4440,3810)
\put(0,4800){$1$}
\put(0,3900){$2$}
\put(0,3000){$3$}
\put(0,1200){$N$}

\put(5600,-200){$1$}
\put(4700,-200){$2$}
\put(2900,-200){$3$}
\put(1100,-200){$N$}

\end{picture}

\begin{ca}
\label{dashedlinesfigure}
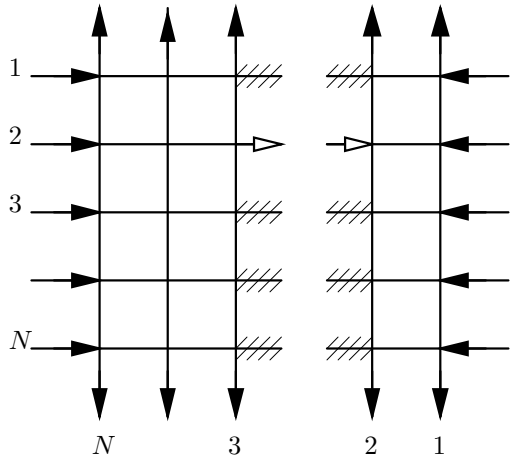
Slicing a domain wall configuration into two parts. The 
white arrows are frozen. The shaded exposed bonds indicate
identified arrows that are summed over.
\end{ca}

\end{minipage}
\end{center}
where the dashed lines in figure \ref{dashedlinesfigure} 
stand for bonds whose arrows are not frozen, and can take 
any orientation that is allowed.

\paragraph*{Exposed arrows} We call the arrows that 
were originally internal, and became external upon slicing 
the lattice, `exposed' arrows. When we slice an
$N$$\times$$N$ lattice into two parts, there will be 
$N$ exposed arrows on each part. The orientations of 
two exposed arrows that originate from slicing an originally 
internal bond are identical, and (generally) summed over.

\paragraph*{Boundary correlation functions} We wish to 
define each part, obtained by slicing a set of domain 
wall configurations, as in figure \ref{dashedlinesfigure}, 
as a boundary correlation function, with the understanding 
that arrows that were frozen by the original dwbc's remain 
as they were, $n$ exposed arrows (on each part) are frozen, 
the rest of the exposed arrows are summed over. In the example 
in figure \ref{dashedlinesfigure}, $n=1$. 

\paragraph*{Simple boundary $n$-point functions} There are 
special cases when {\it apart from} the $n$ frozen arrows, 
the rest of the arrows assume domain wall boundary 
orientations: 
all those on the outside of the vertical boundary lines 
point inwards, and all those on the outside of the horizontal 
boundary lines point inwards. We refer to these as {\it simple}
$n$-point boundary functions. For example, each part in the 
following set of configurations (with all internal arrows 
summed over) is a simple boundary 2-point function:


\begin{center}
\begin{minipage}{4in}
\setlength{\unitlength}{0.001cm}

\begin{picture}(7000,6000)(-1000, 0)

\thicklines

\path(2100,5700)(2100,300)
\path(2100,5025)(2100,5625)
\blacken\path(2190,5265)(2100,5625)(2010,5265)(2190,5265)
\path(2100,900)(2100,300)
\blacken\path(2010,660)(2100,300)(2190,660)(2010,660)
\path(3000,5700)(3000,300)
\path(3000,5100)(3000,5700)
\blacken\path(3090,5340)(3000,5700)(2910,5340)(3090,5340)
\path(3000,900)(3000,300)
\blacken\path(2910,660)(3000,300)(3090,660)(2910,660)
\path(1200,5100)(1200,5700)
\blacken\path(1290,5340)(1200,5700)(1110,5340)(1290,5340)
\path(1200,900)(1200,300)
\blacken\path(1110,660)(1200,300)(1290,660)(1110,660)
\path(1200,5700)(1200,300)
\path(300,2100)(3600,2100)
\path(300,1200)(3600,1200)
\path(300,3000)(3600,3000)
\path(300,3900)(3600,3900)
\path(300,4800)(3600,4800)
\path(4200,4800)(6600,4800)
\path(4200,3900)(6600,3900)
\path(4200,3000)(6600,3000)
\path(4200,1200)(6600,1200)
\path(4800,5700)(4800,300)
\path(5700,5700)(5700,300)
\path(6300,4800)(5700,4800)
\blacken\path(6060,4890)(5700,4800)(6060,4710)(6060,4890)
\path(6300,3900)(5700,3900)
\blacken\path(6060,3990)(5700,3900)(6060,3810)(6060,3990)
\path(6300,3000)(5700,3000)
\blacken\path(6060,3090)(5700,3000)(6060,2910)(6060,3090)
\path(6300,2100)(5700,2100)
\blacken\path(6060,2190)(5700,2100)(6060,2010)(6060,2190)
\path(6300,1200)(5700,1200)
\blacken\path(6060,1290)(5700,1200)(6060,1110)(6060,1290)
\path(600,4800)(1200,4800)
\blacken\path(840,4710)(1200,4800)(840,4890)(840,4710)
\path(600,3900)(1200,3900)
\blacken\path(840,3810)(1200,3900)(840,3990)(840,3810)
\path(600,3000)(1200,3000)
\blacken\path(840,2910)(1200,3000)(840,3090)(840,2910)
\path(600,2100)(1200,2100)
\blacken\path(840,2010)(1200,2100)(840,2190)(840,2010)
\path(600,1200)(1200,1200)
\blacken\path(840,1110)(1200,1200)(840,1290)(840,1110)
\path(4800,5100)(4800,5700)
\blacken\path(4905,5340)(4800,5700)(4695,5340)(4905,5340)
\path(5700,5100)(5700,5700)
\blacken\path(5805,5340)(5700,5700)(5595,5340)(5805,5340)
\path(4800,900)(4800,300)
\blacken\path(4710,660)(4800,300)(4890,660)(4710,660)
\path(5700,900)(5700,300)
\blacken\path(5610,660)(5700,300)(5790,660)(5610,660)
\path(4800,4800)(4200,4800)
\blacken\path(4560,4890)(4200,4800)(4560,4710)(4560,4890)
\path(3525,4800)(2925,4800)
\blacken\path(3285,4890)(2925,4800)(3285,4710)(3285,4890)
\path(3000,3900)(3600,3900)
\whiten\path(3240,3810)(3600,3900)(3240,3990)(3240,3810)
\path(4200,3900)(4800,3900)
\whiten\path(4440,3810)(4800,3900)(4440,3990)(4440,3810)
\path(4800,3000)(4200,3000)
\blacken\path(4560,3090)(4200,3000)(4560,2910)(4560,3090)
\path(3600,3000)(3000,3000)
\blacken\path(3360,3090)(3000,3000)(3360,2910)(3360,3090)
\path(4200,2100)(6600,2100)
\path(3000,2100)(3600,2100)
\whiten\path(3240,2010)(3600,2100)(3240,2190)(3240,2010)
\path(4200,2100)(4800,2100)
\whiten\path(4440,2010)(4800,2100)(4440,2190)(4440,2010)
\path(3525,1200)(2925,1200)
\blacken\path(3285,1290)(2925,1200)(3285,1110)(3285,1290)
\path(4800,1200)(4200,1200)
\blacken\path(4560,1290)(4200,1200)(4560,1110)(4560,1290)

\put(0,4800){$1$}
\put(0,3900){$2$}
\put(0,3000){$3$}
\put(0,1200){$N$}

\put(5600,-200){$1$}
\put(4700,-200){$2$}
\put(2900,-200){$3$}
\put(1100,-200){$N$}

\end{picture}

\begin{ca}
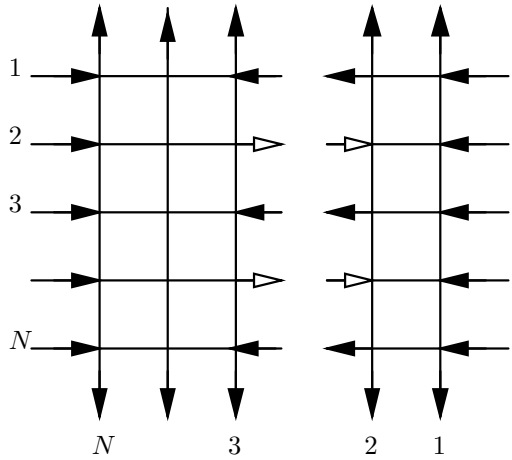
Two parts, each of which is a simple boundary 2-point function.
\end{ca}
\end{minipage}
\end{center}

\paragraph*{Origin of the simple boundary functions}

\phantom{{}}From dwbc's, and conservation of arrow flow, there 
is exactly one horizontal arrow, pointing to the right, on the 
left of the right boundary, two such arrows to left of the 
second column, three to the left of the third, and so forth, 
until we reach the left boundary and find the correct dwbc's.


\begin{center}
\begin{minipage}{4in}
\setlength{\unitlength}{0.001cm}

\begin{picture}(6000,6000)(-2000, 0)
\thicklines

\path(300,4800)(5700,4800)
\path(300,3900)(5700,3900)
\path(300,2100)(5700,2100)
\path(300,1200)(5700,1200)
\path(1200,5700)(1200,300)
\path(2100,5700)(2100,300)
\path(4800,5700)(4800,300)
\path(1200,5025)(1200,5625)
\blacken\path(1290,5265)(1200,5625)(1110,5265)(1290,5265)
\path(2100,5025)(2100,5625)
\blacken\path(2190,5265)(2100,5625)(2010,5265)(2190,5265)
\path(3900,5025)(3900,5625)
\blacken\path(3990,5265)(3900,5625)(3810,5265)(3990,5265)
\path(4800,5100)(4800,5700)
\blacken\path(4905,5340)(4800,5700)(4695,5340)(4905,5340)
\path(1425,4800)(2025,4800)
\whiten\path(1665,4710)(2025,4800)(1665,4890)(1665,4710)
\path(3675,4800)(3075,4800)
\blacken\path(3435,4890)(3075,4800)(3435,4710)(3435,4890)
\path(4500,4800)(3900,4800)
\blacken\path(4260,4890)(3900,4800)(4260,4710)(4260,4890)
\path(5400,4800)(4800,4800)
\blacken\path(5160,4890)(4800,4800)(5160,4710)(5160,4890)
\path(525,3900)(1125,3900)
\whiten\path(765,3810)(1125,3900)(765,3990)(765,3810)
\path(1800,3900)(1200,3900)
\blacken\path(1560,3990)(1200,3900)(1560,3810)(1560,3990)
\path(5400,3900)(4800,3900)
\blacken\path(5160,3990)(4800,3900)(5160,3810)(5160,3990)
\path(600,2100)(1200,2100)
\whiten\path(840,2010)(1200,2100)(840,2190)(840,2010)
\path(600,1200)(1200,1200)
\whiten\path(840,1110)(1200,1200)(840,1290)(840,1110)
\path(3600,3000)(3000,3000)
\blacken\path(3360,3090)(3000,3000)(3360,2910)(3360,3090)
\path(4575,3000)(3975,3000)
\blacken\path(4335,3090)(3975,3000)(4335,2910)(4335,3090)
\path(5475,3000)(4875,3000)
\blacken\path(5235,3090)(4875,3000)(5235,2910)(5235,3090)
\path(5400,2100)(4800,2100)
\blacken\path(5160,2190)(4800,2100)(5160,2010)(5160,2190)
\path(5475,1200)(4875,1200)
\blacken\path(5235,1290)(4875,1200)(5235,1110)(5235,1290)
\path(4800,1800)(4800,1200)
\blacken\path(4710,1560)(4800,1200)(4890,1560)(4710,1560)
\path(1500,1200)(2100,1200)
\whiten\path(1740,1110)(2100,1200)(1740,1290)(1740,1110)
\path(4575,1200)(3975,1200)
\blacken\path(4335,1290)(3975,1200)(4335,1110)(4335,1290)
\path(1200,975)(1200,375)
\blacken\path(1110,735)(1200,375)(1290,735)(1110,735)
\path(2100,900)(2100,300)
\blacken\path(2010,660)(2100,300)(2190,660)(2010,660)
\path(3900,900)(3900,300)
\blacken\path(3810,660)(3900,300)(3990,660)(3810,660)
\path(4800,900)(4800,300)
\blacken\path(4710,660)(4800,300)(4890,660)(4710,660)
\path(525,4800)(1125,4800)
\whiten\path(765,4710)(1125,4800)(765,4890)(765,4710)
\path(4200,2100)(4800,2100)
\whiten\path(4440,2010)(4800,2100)(4440,2190)(4440,2010)
\path(4800,2400)(4800,3000)
\blacken\path(4890,2640)(4800,3000)(4710,2640)(4890,2640)
\path(4800,3375)(4800,3975)
\blacken\path(4890,3615)(4800,3975)(4710,3615)(4890,3615)
\path(4800,4200)(4800,4800)
\blacken\path(4890,4440)(4800,4800)(4710,4440)(4890,4440)
\path(4500,3900)(3900,3900)
\blacken\path(4260,3990)(3900,3900)(4260,3810)(4260,3990)
\path(3000,5700)(3000,300)
\path(3900,4200)(3900,4800)
\blacken\path(3990,4440)(3900,4800)(3810,4440)(3990,4440)
\path(3900,5700)(3900,300)
\path(3300,3900)(3900,3900)
\whiten\path(3540,3810)(3900,3900)(3540,3990)(3540,3810)
\path(3900,3600)(3900,3000)
\blacken\path(3810,3360)(3900,3000)(3990,3360)(3810,3360)
\path(3900,2700)(3900,2100)
\blacken\path(3810,2460)(3900,2100)(3990,2460)(3810,2460)
\path(3000,5100)(3000,5700)
\blacken\path(3090,5340)(3000,5700)(2910,5340)(3090,5340)
\path(3000,4500)(3000,3900)
\blacken\path(2910,4260)(3000,3900)(3090,4260)(2910,4260)
\path(3000,3300)(3000,3900)
\blacken\path(3090,3540)(3000,3900)(2910,3540)(3090,3540)
\path(3000,2400)(3000,3000)
\blacken\path(3090,2640)(3000,3000)(2910,2640)(3090,2640)
\path(3000,900)(3000,300)
\blacken\path(2910,660)(3000,300)(3090,660)(2910,660)
\path(2100,2700)(2100,2100)
\blacken\path(2010,2460)(2100,2100)(2190,2460)(2010,2460)
\path(2100,3375)(2100,3975)
\blacken\path(2190,3615)(2100,3975)(2010,3615)(2190,3615)
\path(2100,4200)(2100,4800)
\blacken\path(2190,4440)(2100,4800)(2010,4440)(2190,4440)
\path(1200,4275)(1200,4875)
\blacken\path(1290,4515)(1200,4875)(1110,4515)(1290,4515)
\path(1200,3600)(1200,3000)
\blacken\path(1110,3360)(1200,3000)(1290,3360)(1110,3360)
\path(1200,2700)(1200,2100)
\blacken\path(1110,2460)(1200,2100)(1290,2460)(1110,2460)
\path(1200,1800)(1200,1200)
\blacken\path(1110,1560)(1200,1200)(1290,1560)(1110,1560)
\path(3300,2100)(3900,2100)
\whiten\path(3540,2010)(3900,2100)(3540,2190)(3540,2010)
\path(3900,1800)(3900,1200)
\blacken\path(3810,1560)(3900,1200)(3990,1560)(3810,1560)
\path(3600,1200)(3000,1200)
\blacken\path(3360,1290)(3000,1200)(3360,1110)(3360,1290)
\path(2400,4800)(3000,4800)
\whiten\path(2640,4710)(3000,4800)(2640,4890)(2640,4710)
\path(2625,3900)(2025,3900)
\blacken\path(2385,3990)(2025,3900)(2385,3810)(2385,3990)
\path(2700,3000)(2100,3000)
\blacken\path(2460,3090)(2100,3000)(2460,2910)(2460,3090)
\path(3000,1500)(3000,2100)
\blacken\path(3090,1740)(3000,2100)(2910,1740)(3090,1740)
\path(2400,2100)(3000,2100)
\whiten\path(2640,2010)(3000,2100)(2640,2190)(2640,2010)
\path(2400,1200)(3000,1200)
\whiten\path(2640,1110)(3000,1200)(2640,1290)(2640,1110)
\path(1500,2100)(2100,2100)
\whiten\path(1740,2010)(2100,2100)(1740,2190)(1740,2010)
\path(2100,1800)(2100,1200)
\blacken\path(2010,1560)(2100,1200)(2190,1560)(2010,1560)
\path(300,3000)(5700,3000)
\path(1500,3000)(2100,3000)
\whiten\path(1740,2910)(2100,3000)(1740,3090)(1740,2910)
\path(600,3000)(1200,3000)
\whiten\path(840,2910)(1200,3000)(840,3090)(840,2910)

\put(0,4800){$1$}
\put(0,3900){$2$}
\put(0,3000){$3$}
\put(0,1200){$N$}

\put(4700,-200){$1$}
\put(3800,-200){$2$}
\put(2900,-200){$3$}
\put(1100,-200){$N$}

\end{picture}

\begin{ca}
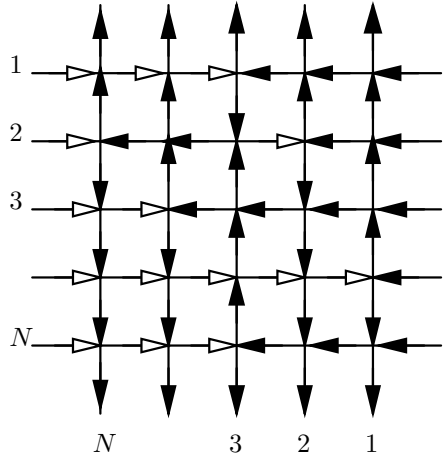
Scanning the white arrows from right to left, we can see 
how the number of horizontal arrows pointing to the right 
increases by 1, every time we step to the left. 
\end{ca}
\end{minipage}
\end{center}

Simple $n$-point functions, with $n$ horizontal arrows, all 
on the same column, are obtained by slicing a domain wall 
configuration vertically, between the $n$-th and vertical 
line (counting from the right) and the ($n$$+$$1$)-st line. 

Precisely the same arguments apply to horizontal arrows that 
point to the left, and vertical arrows that point either 
up or down.

\paragraph*{Composite boundary $n$-point functions} are 
obtained, for example, when an $N$$\times$$N$ lattice is sliced, 
$N_{right}$ exposed arrows are pointing to the right, 
$N - N_{right}$ are pointing to left, and we freeze $n$ arrows 
that point, let's say, to the right. If $n = N_{right}$, then 
the boundary function is simple. If $n < N_{right}$, we need 
to sum over the positions of the remaining right-pointing arrows. 
Hence, such a boundary function is not simple.

\paragraph*{Bulk correlation functions}

In principle, bulk $n$-point function, where all arrows point 
in the same direction, and lie on the same line, are obtained 
by taking products of boundary functions corresponding to each 
part.

More general bulk functions can be obtained by taking products
of more general boundary functions. We have nothing deep to say 
about bulk functions at this stage.

\section{Simple boundary 1-point functions}

The simplest correlation function to compute is a simple 
boundary 1-point function. This will correspond to freezing 
any one of the white arrows shown in figure
\ref{invertedsinglearrowsfigure}


\begin{center}
\begin{minipage}{4in}
\setlength{\unitlength}{0.001cm}

\begin{picture}(6000,6000)(-1500, 0)

\thicklines

\path(300,4800)(5700,4800)
\path(300,3900)(5700,3900)
\path(300,2100)(5700,2100)
\path(300,1200)(5700,1200)
\path(1200,5700)(1200,300)
\path(2100,5700)(2100,300)
\path(4800,5700)(4800,300)
\path(1200,5025)(1200,5625)
\blacken\path(1290,5265)(1200,5625)(1110,5265)(1290,5265)
\path(2100,5025)(2100,5625)
\blacken\path(2190,5265)(2100,5625)(2010,5265)(2190,5265)
\path(3900,5025)(3900,5625)
\blacken\path(3990,5265)(3900,5625)(3810,5265)(3990,5265)
\path(4800,5100)(4800,5700)
\blacken\path(4905,5340)(4800,5700)(4695,5340)(4905,5340)
\path(1425,4800)(2025,4800)
\blacken\path(1665,4710)(2025,4800)(1665,4890)(1665,4710)
\path(3675,4800)(3075,4800)
\blacken\path(3435,4890)(3075,4800)(3435,4710)(3435,4890)
\path(4500,4800)(3900,4800)
\blacken\path(4260,4890)(3900,4800)(4260,4710)(4260,4890)
\path(5400,4800)(4800,4800)
\blacken\path(5160,4890)(4800,4800)(5160,4710)(5160,4890)
\path(525,3900)(1125,3900)
\blacken\path(765,3810)(1125,3900)(765,3990)(765,3810)
\path(1800,3900)(1200,3900)
\whiten\path(1560,3990)(1200,3900)(1560,3810)(1560,3990)
\path(5400,3900)(4800,3900)
\blacken\path(5160,3990)(4800,3900)(5160,3810)(5160,3990)
\path(600,2100)(1200,2100)
\blacken\path(840,2010)(1200,2100)(840,2190)(840,2010)
\path(600,1200)(1200,1200)
\blacken\path(840,1110)(1200,1200)(840,1290)(840,1110)
\path(3600,3000)(3000,3000)
\blacken\path(3360,3090)(3000,3000)(3360,2910)(3360,3090)
\path(4575,3000)(3975,3000)
\blacken\path(4335,3090)(3975,3000)(4335,2910)(4335,3090)
\path(5475,3000)(4875,3000)
\blacken\path(5235,3090)(4875,3000)(5235,2910)(5235,3090)
\path(5400,2100)(4800,2100)
\blacken\path(5160,2190)(4800,2100)(5160,2010)(5160,2190)
\path(5475,1200)(4875,1200)
\blacken\path(5235,1290)(4875,1200)(5235,1110)(5235,1290)
\path(4800,1800)(4800,1200)
\blacken\path(4710,1560)(4800,1200)(4890,1560)(4710,1560)
\path(1500,1200)(2100,1200)
\blacken\path(1740,1110)(2100,1200)(1740,1290)(1740,1110)
\path(4575,1200)(3975,1200)
\blacken\path(4335,1290)(3975,1200)(4335,1110)(4335,1290)
\path(1200,975)(1200,375)
\blacken\path(1110,735)(1200,375)(1290,735)(1110,735)
\path(2100,900)(2100,300)
\blacken\path(2010,660)(2100,300)(2190,660)(2010,660)
\path(3900,900)(3900,300)
\blacken\path(3810,660)(3900,300)(3990,660)(3810,660)
\path(4800,900)(4800,300)
\blacken\path(4710,660)(4800,300)(4890,660)(4710,660)
\path(525,4800)(1125,4800)
\blacken\path(765,4710)(1125,4800)(765,4890)(765,4710)
\path(4200,2100)(4800,2100)
\whiten\path(4440,2010)(4800,2100)(4440,2190)(4440,2010)
\path(4800,2400)(4800,3000)
\blacken\path(4890,2640)(4800,3000)(4710,2640)(4890,2640)
\path(4800,3375)(4800,3975)
\blacken\path(4890,3615)(4800,3975)(4710,3615)(4890,3615)
\path(4800,4200)(4800,4800)
\blacken\path(4890,4440)(4800,4800)(4710,4440)(4890,4440)
\path(4500,3900)(3900,3900)
\blacken\path(4260,3990)(3900,3900)(4260,3810)(4260,3990)
\path(3000,5700)(3000,300)
\path(3900,4200)(3900,4800)
\blacken\path(3990,4440)(3900,4800)(3810,4440)(3990,4440)
\path(3900,5700)(3900,300)
\path(3300,3900)(3900,3900)
\blacken\path(3540,3810)(3900,3900)(3540,3990)(3540,3810)
\path(3900,3600)(3900,3000)
\blacken\path(3810,3360)(3900,3000)(3990,3360)(3810,3360)
\path(3900,2700)(3900,2100)
\blacken\path(3810,2460)(3900,2100)(3990,2460)(3810,2460)
\path(3000,5100)(3000,5700)
\blacken\path(3090,5340)(3000,5700)(2910,5340)(3090,5340)
\path(3000,4500)(3000,3900)
\whiten\path(2910,4260)(3000,3900)(3090,4260)(2910,4260)
\path(3000,3300)(3000,3900)
\blacken\path(3090,3540)(3000,3900)(2910,3540)(3090,3540)
\path(3000,2400)(3000,3000)
\blacken\path(3090,2640)(3000,3000)(2910,2640)(3090,2640)
\path(3000,900)(3000,300)
\blacken\path(2910,660)(3000,300)(3090,660)(2910,660)
\path(2100,2700)(2100,2100)
\blacken\path(2010,2460)(2100,2100)(2190,2460)(2010,2460)
\path(2100,3375)(2100,3975)
\blacken\path(2190,3615)(2100,3975)(2010,3615)(2190,3615)
\path(2100,4200)(2100,4800)
\blacken\path(2190,4440)(2100,4800)(2010,4440)(2190,4440)
\path(1200,4275)(1200,4875)
\blacken\path(1290,4515)(1200,4875)(1110,4515)(1290,4515)
\path(1200,3600)(1200,3000)
\blacken\path(1110,3360)(1200,3000)(1290,3360)(1110,3360)
\path(1200,2700)(1200,2100)
\blacken\path(1110,2460)(1200,2100)(1290,2460)(1110,2460)
\path(1200,1800)(1200,1200)
\blacken\path(1110,1560)(1200,1200)(1290,1560)(1110,1560)
\path(3300,2100)(3900,2100)
\blacken\path(3540,2010)(3900,2100)(3540,2190)(3540,2010)
\path(3900,1800)(3900,1200)
\blacken\path(3810,1560)(3900,1200)(3990,1560)(3810,1560)
\path(3600,1200)(3000,1200)
\blacken\path(3360,1290)(3000,1200)(3360,1110)(3360,1290)
\path(2400,4800)(3000,4800)
\blacken\path(2640,4710)(3000,4800)(2640,4890)(2640,4710)
\path(2625,3900)(2025,3900)
\blacken\path(2385,3990)(2025,3900)(2385,3810)(2385,3990)
\path(2700,3000)(2100,3000)
\blacken\path(2460,3090)(2100,3000)(2460,2910)(2460,3090)
\path(3000,1500)(3000,2100)
\whiten\path(3090,1740)(3000,2100)(2910,1740)(3090,1740)
\path(2400,2100)(3000,2100)
\blacken\path(2640,2010)(3000,2100)(2640,2190)(2640,2010)
\path(2400,1200)(3000,1200)
\blacken\path(2640,1110)(3000,1200)(2640,1290)(2640,1110)
\path(1500,2100)(2100,2100)
\blacken\path(1740,2010)(2100,2100)(1740,2190)(1740,2010)
\path(2100,1800)(2100,1200)
\blacken\path(2010,1560)(2100,1200)(2190,1560)(2010,1560)
\path(300,3000)(5700,3000)
\path(1500,3000)(2100,3000)
\blacken\path(1740,2910)(2100,3000)(1740,3090)(1740,2910)
\path(600,3000)(1200,3000)
\blacken\path(840,2910)(1200,3000)(840,3090)(840,2910)

\put(-100,4800){1}
\put(-100,3900){2}
\put(-100,3000){3}
\put(-100,1200){$N$}

\put(4700,-200){1}
\put(3800,-200){2}
\put(2900,-200){3}
\put(1100,-200){$N$}

\end{picture}

\begin{ca}
\label{invertedsinglearrowsfigure}
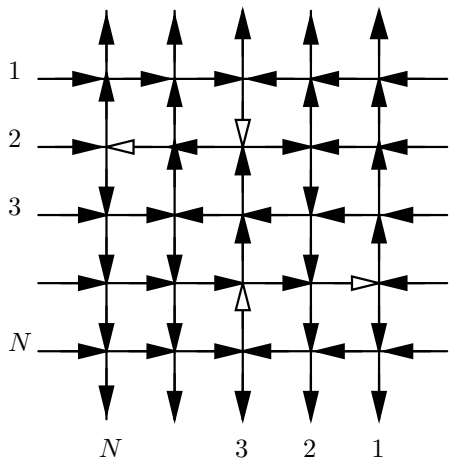
There is exactly one internal arrow, that touches a boundary, 
and that points opposite to the external arrows on the other 
side of that boundary. 
\end{ca}

\end{minipage}
\end{center}

We choose to freeze the horizontal right-oriented 
arrow. Using the algebraic Bethe {\it ansatz}, Bogoliubov 
{\it et al} \cite{BPZ} obtained a determinant representation 
of this correlation function. More precisely, they 
obtained the partition function of the set of all domain 
wall boundary configurations such that the $c$ vertex on 
the right boundary lies on row $r$, normalized
by the full partition function $Z_N$. This is their definition
of a boundary 1-point boundary function. 

To reproduce their result, we start by slicing the lattice 
vertically, into two parts, and expose the frozen arrow


\begin{center}
\begin{minipage}{4in}
\setlength{\unitlength}{0.001cm}

\begin{picture}(6000,6000)(-2000, 0)
\thicklines

\path(1200,5700)(1200,300)
\path(2100,5700)(2100,300)
\path(3000,5700)(3000,300)
\path(3900,5700)(3900,300)
\path(300,1200)(4500,1200)
\path(300,2100)(4500,2100)
\path(300,3000)(4500,3000)
\path(225,3900)(4425,3900)
\path(300,4800)(4500,4800)
\path(4800,4800)(6000,4800)
\path(4800,3900)(6000,3900)
\path(4800,3000)(6000,3000)
\path(4800,2100)(6000,2100)
\path(4800,1200)(6000,1200)
\path(5400,5700)(5400,300)
\path(1200,5100)(1200,5700)
\blacken\path(1290,5340)(1200,5700)(1110,5340)(1290,5340)
\path(2100,5100)(2100,5700)
\blacken\path(2190,5340)(2100,5700)(2010,5340)(2190,5340)
\path(3000,5100)(3000,5700)
\blacken\path(3090,5340)(3000,5700)(2910,5340)(3090,5340)
\path(3900,5100)(3900,5700)
\blacken\path(3990,5340)(3900,5700)(3810,5340)(3990,5340)
\path(5400,5100)(5400,5700)
\blacken\path(5490,5340)(5400,5700)(5310,5340)(5490,5340)
\path(6000,4800)(5400,4800)
\blacken\path(5760,4890)(5400,4800)(5760,4710)(5760,4890)
\path(6000,3000)(5400,3000)
\blacken\path(5760,3090)(5400,3000)(5760,2910)(5760,3090)
\path(6000,2100)(5400,2100)
\blacken\path(5760,2190)(5400,2100)(5760,2010)(5760,2190)
\path(6000,1200)(5400,1200)
\blacken\path(5760,1290)(5400,1200)(5760,1110)(5760,1290)
\path(5400,1200)(4800,1200)
\blacken\path(5160,1290)(4800,1200)(5160,1110)(5160,1290)
\path(5400,3000)(4800,3000)
\blacken\path(5160,3090)(4800,3000)(5160,2910)(5160,3090)
\path(5400,3900)(4800,3900)
\blacken\path(5160,3990)(4800,3900)(5160,3810)(5160,3990)
\path(5400,4800)(4800,4800)
\blacken\path(5160,4890)(4800,4800)(5160,4710)(5160,4890)
\path(4500,4800)(3900,4800)
\blacken\path(4260,4890)(3900,4800)(4260,4710)(4260,4890)
\path(4500,3900)(3900,3900)
\blacken\path(4260,3990)(3900,3900)(4260,3810)(4260,3990)
\path(4500,3000)(3900,3000)
\blacken\path(4260,3090)(3900,3000)(4260,2910)(4260,3090)
\path(4500,1200)(3900,1200)
\blacken\path(4260,1290)(3900,1200)(4260,1110)(4260,1290)
\path(6000,3900)(5400,3900)
\blacken\path(5760,3990)(5400,3900)(5760,3810)(5760,3990)
\path(525,4800)(1125,4800)
\blacken\path(765,4710)(1125,4800)(765,4890)(765,4710)
\path(525,3900)(1125,3900)
\blacken\path(765,3810)(1125,3900)(765,3990)(765,3810)
\path(525,3000)(1125,3000)
\blacken\path(765,2910)(1125,3000)(765,3090)(765,2910)
\path(525,2100)(1125,2100)
\blacken\path(765,2010)(1125,2100)(765,2190)(765,2010)
\path(600,1200)(1200,1200)
\blacken\path(840,1110)(1200,1200)(840,1290)(840,1110)
\path(3900,2100)(4500,2100)
\whiten\path(4140,2010)(4500,2100)(4140,2190)(4140,2010)
\path(4800,2100)(5400,2100)
\whiten\path(5040,2010)(5400,2100)(5040,2190)(5040,2010)
\path(1200,900)(1200,300)
\blacken\path(1110,660)(1200,300)(1290,660)(1110,660)
\path(2100,900)(2100,300)
\blacken\path(2010,660)(2100,300)(2190,660)(2010,660)
\path(3000,900)(3000,300)
\blacken\path(2910,660)(3000,300)(3090,660)(2910,660)
\path(3900,900)(3900,300)
\blacken\path(3810,660)(3900,300)(3990,660)(3810,660)
\path(5400,900)(5400,300)
\blacken\path(5310,660)(5400,300)(5490,660)(5310,660)
\path(5400,4200)(5400,4800)
\blacken\path(5490,4440)(5400,4800)(5310,4440)(5490,4440)
\path(5400,3300)(5400,3900)
\blacken\path(5490,3540)(5400,3900)(5310,3540)(5490,3540)
\path(5400,2400)(5400,3000)
\blacken\path(5490,2640)(5400,3000)(5310,2640)(5490,2640)
\path(5400,1800)(5400,1200)
\blacken\path(5310,1560)(5400,1200)(5490,1560)(5310,1560)

\put(0,4800){$1$}
\put(0,3900){$2$}
\put(0,3000){$3$}
\put(0,1200){$N$}
\put(5300,-200){$1$}
\put(3800,-200){$2$}
\put(2900,-200){$3$}
\put(1100,-200){$N$}

\end{picture}

\begin{ca}
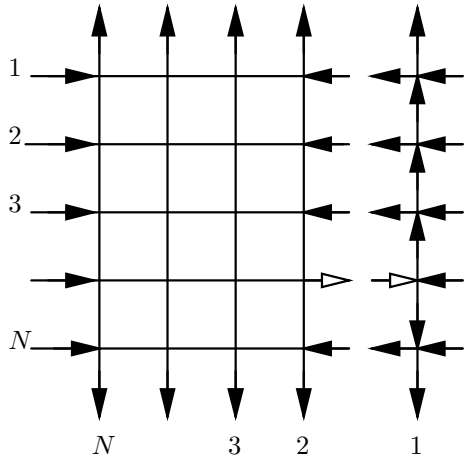
Slicing a domain wall configuration into two 1-point functions.
\end{ca}

\end{minipage}
\end{center}

We end up with {\it two} simple 1-point functions: One on the
right, and one on the left. The right 1-point function is 
can be computed by inspection. The left will 
be computed below. The result of Bogoliubov {\it et al} is 
the product of the two parts, normalized by the partition 
function.

\subsection{The right boundary 1-point function}

Consider the partition function of the $1$$\times$$N$ 
right part, and take the single $c$ vertex to be on row 
$r$. 

Given the dwbc's and conservation of arrow flows, all 
vertices that are above row $r$, will be of type $b$, 
all those below will be of type $a$, and the weight of the 
right part is

\begin{equation}
\label{rightpartequation}
\ll \prod_{i=1  }^{r-1} b(x_r, y_1) \rr
                                 c(x_r,y_1)
\ll \prod_{i=r+1}^{N  } a(x_i,y_1) \rr
\end{equation}

\subsection{The left boundary 1-point function}

Consider the partition function of the ($N$$-$$1$)$\times$$N$
left part. If the inverted arrow on the right boundary were 
all the way at the top row, we would have the situation shown 
in figure \ref{separatingrowfigure}.


\begin{center}
\begin{minipage}{3in}
\setlength{\unitlength}{0.001cm}

\begin{picture}(6000,6000)(-1000, 0)

\thicklines

\path(1822,4500)(1822,250)
\path(2722,4500)(2722,250)
\path(3622,4500)(3622,250)
\path(22,3900)(4222,3900)
\path(22,3000)(4222,3000)
\path(22,2100)(4222,2100)
\path(22,1200)(4222,1200)
\path(322,3900)(922,3900)
\blacken\path(562,3810)(922,3900)(562,3990)(562,3810)
\path(322,3000)(922,3000)
\blacken\path(562,2910)(922,3000)(562,3090)(562,2910)
\path(322,2100)(922,2100)
\blacken\path(562,2010)(922,2100)(562,2190)(562,2010)
\path(322,1200)(922,1200)
\blacken\path(562,1110)(922,1200)(562,1290)(562,1110)
\path(922,900)(922,300)
\blacken\path(832,660)(922,300)(1012,660)(832,660)
\path(1822,900)(1822,300)
\blacken\path(1732,660)(1822,300)(1912,660)(1732,660)
\path(2722,900)(2722,300)
\blacken\path(2632,660)(2722,300)(2812,660)(2632,660)
\path(3622,900)(3622,300)
\blacken\path(3532,660)(3622,300)(3712,660)(3532,660)
\path(4222,3900)(3622,3900)
\blacken\path(3982,3990)(3622,3900)(3982,3810)(3982,3990)
\path(4222,3000)(3622,3000)
\blacken\path(3982,3090)(3622,3000)(3982,2910)(3982,3090)
\path(4222,1200)(3622,1200)
\blacken\path(3982,1290)(3622,1200)(3982,1110)(3982,1290)
\path(4222,2100)(3622,2100)
\blacken\path(3982,2190)(3622,2100)(3982,2010)(3982,2190)
\path(922,4500)(922,250)
\path(22,5400)(4222,5400)
\path(922,6000)(922,4800)
\path(1822,6000)(1822,4800)
\path(2722,6000)(2722,4800)
\path(3622,6000)(3622,4800)
\path(322,5400)(922,5400)
\blacken\path(562,5310)(922,5400)(562,5490)(562,5310)
\path(922,5400)(922,6000)
\blacken\path(1012,5640)(922,6000)(832,5640)(1012,5640)
\path(1822,5400)(1822,6000)
\blacken\path(1912,5640)(1822,6000)(1732,5640)(1912,5640)
\path(2722,5400)(2722,6000)
\blacken\path(2812,5640)(2722,6000)(2632,5640)(2812,5640)
\path(3622,5400)(3622,6000)
\blacken\path(3712,5640)(3622,6000)(3532,5640)(3712,5640)
\path(922,4800)(922,5400)
\blacken\path(1012,5040)(922,5400)(832,5040)(1012,5040)
\path(1822,4800)(1822,5400)
\blacken\path(1912,5040)(1822,5400)(1732,5040)(1912,5040)
\path(2722,4800)(2722,5400)
\blacken\path(2812,5040)(2722,5400)(2632,5040)(2812,5040)
\path(3622,4800)(3622,5400)
\blacken\path(3712,5040)(3622,5400)(3532,5040)(3712,5040)
\path(922,3900)(922,4500)
\blacken\path(1012,4140)(922,4500)(832,4140)(1012,4140)
\path(1822,3900)(1822,4500)
\blacken\path(1912,4140)(1822,4500)(1732,4140)(1912,4140)
\path(2722,3900)(2722,4500)
\blacken\path(2812,4140)(2722,4500)(2632,4140)(2812,4140)
\path(3622,3900)(3622,4500)
\blacken\path(3712,4140)(3622,4500)(3532,4140)(3712,4140)
\path(1222,5400)(1822,5400)
\blacken\path(1462,5310)(1822,5400)(1462,5490)(1462,5310)
\path(2122,5400)(2722,5400)
\blacken\path(2362,5310)(2722,5400)(2362,5490)(2362,5310)
\path(3022,5400)(3622,5400)
\blacken\path(3262,5310)(3622,5400)(3262,5490)(3262,5310)
\path(3622,5400)(4222,5400)
\whiten\path(3862,5310)(4222,5400)(3862,5490)(3862,5310)

\put(0847,-100){$N$}
\put(2647,-100){3}
\put(3547,-100){2}

\end{picture}

\begin{ca}
\label{separatingrowfigure}
`Peeling' a frozen row.
\end{ca}

\end{minipage}
\end{center}

In this case, all vertices on the top row are frozen to be 
of type $a$. Therefore, we can peel that top row, and end 
up with configurations on an ($N$$-$$1$)$\times$($N$$-$$1$) 
square lattice that satisfy dwbc's, and whose partition 
function can be evaluated using Izergin's expression.

However, the inverted arrow is, in general, not at the top 
row. To bring it to the top row, we use the Yang Baxter 
equation as follows.

\subsection{Rolling once}

Consider the set of all configurations that remain after 
peeling the right boundary. They live on an
($N$$-$$1$)$\times$$N$ lattice, with dwbc's, except for 
an inverted arrow at row $r$, on the right boundary. 

Take the inverted arrow to be (initially) at row $i$, 
and take the horizontal rapidity that flows through it 
to be (initially) $x_{i}$. We denote the partition 
function of the set of all such configurations by 
$F[r_i, x_i]$, and represent them schematically by 
the graph on the left hand side of figure \ref{addingtwofigure}. 

Consider another set of configurations, 
$F[r_{i-1}, x_{i-1}]$, that is identical to 
$F[r_{i  }, x_{i  }]$, except that 
the inverted arrow is now on row $r_{i-1}$, 
and represent it schematically as the graph on the 
right in figure \ref{addingtwofigure}. 


\begin{center}
\begin{minipage}{4in}
\setlength{\unitlength}{0.001cm}
\begin{picture}(8000,3000)(-500, 0)

\thicklines
\path(1350,2722)(1350,22)
\path(2250,2722)(2250,22)
\blacken\path(1440,2362)(1350,2722)(1260,2362)(1440,2362)
\path(1350,2722)(1350,2272)
\blacken\path(1260,382)(1350,22)(1440,382)(1260,382)
\path(1350,22)(1350,472)
\blacken\path(2160,382)(2250,22)(2340,382)(2160,382)
\path(2250,22)(2250,472)
\blacken\path(2340,2362)(2250,2722)(2160,2362)(2340,2362)
\path(2250,2722)(2250,2272)
\thinlines
\path(1650,2422)(1350,2122)
\path(1650,2422)(1350,2122)
\path(1950,2422)(1350,1822)
\path(1950,2422)(1350,1822)
\path(2250,2422)(1350,1522)
\path(2250,2422)(1350,1522)
\path(2250,2122)(1350,1222)
\path(2250,2122)(1350,1222)
\path(2250,1822)(1350,922)
\path(2250,1822)(1350,922)
\path(2250,1522)(1350,622)
\path(2250,1522)(1350,622)
\path(2250,1222)(1350,322)
\path(2250,1222)(1350,322)
\path(2250,922)(1650,322)
\path(2250,922)(1650,322)
\path(2250,622)(1950,322)
\path(2250,622)(1950,322)
\thicklines
\path(750,1822)(1350,1822)
\blacken\path(990,1732)(1350,1822)(990,1912)(990,1732)
\blacken\path(2610,1912)(2250,1822)(2610,1732)(2610,1912)
\path(2250,1822)(2850,1822)
\path(750,922)(1350,922)
\blacken\path(990,832)(1350,922)(990,1012)(990,832)
\path(2250,922)(2850,922)

\path(1350,1822)(2250,1822)
\path(1350,922)(2250,922)
\blacken\path(6540,2362)(6450,2722)(6360,2362)(6540,2362)
\path(6450,2722)(6450,22)
\blacken\path(6360,382)(6450,22)(6540,382)(6360,382)
\blacken\path(7440,2362)(7350,2722)(7260,2362)(7440,2362)
\path(7350,2722)(7350,22)
\blacken\path(7260,382)(7350,22)(7440,382)(7260,382)
\path(5850,1822)(6450,1822)
\blacken\path(6090,1732)(6450,1822)(6090,1912)(6090,1732)
\path(5850,922)(6450,922)
\blacken\path(6090,832)(6450,922)(6090,1012)(6090,832)
\path(7350,1822)(7950,1822)
\blacken\path(7710,1012)(7350,922)(7710,832)(7710,1012)
\path(7350,922)(7950,922)
\thinlines
\path(6750,2422)(6450,2122)
\path(7050,2422)(6450,1822)
\path(7350,2422)(6450,1522)
\path(7350,2122)(6450,1222)
\path(7350,1822)(6450,922)
\path(7350,1522)(6450,622)
\path(7350,1222)(6450,322)
\path(7350,922)(6750,322)
\path(7350,622)(7050,322)
\thicklines
\path(6450,1822)(7350,1822)
\path(6450,0922)(7350,922)

\put(0000,1822){$x_{i-1}$}
\put(0000,0922){$x_{i  }$}

\put(5000,1822){$x_{i-1}$}
\put(5000,0922){$x_{i  }$}

\whiten\path(2490,832)(2850,922)(2490,1012)(2490,832)
\whiten\path(7590,1732)(7950,1822)(7590,1912)(7590,1732)

\end{picture}

\begin{ca}
\label{addingtwofigure}
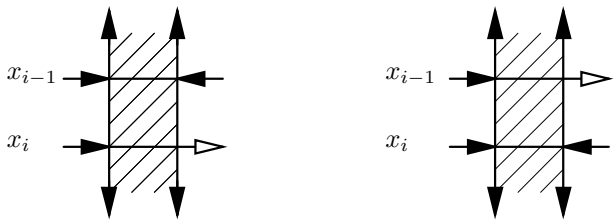
The graph on the left is a schematic presentation of 
the set of configurations with an inverted arrow that 
we wish to roll upwards. The figure on the left is 
an identical set of configurations, apart from the 
fact that the inverted arrow is one row higher.
\end{ca}

\end{minipage}
\end{center}

Multiplying the graph on the left  by $b(x_{i-1}, x_i)$,
and         the graph on the right by $c(x_{i-1}, x_i)$,
and adding the results


\begin{center}
\begin{minipage}{4in}
\setlength{\unitlength}{0.001cm}

\begin{picture}(10000,3000)(-750, 0)

\thicklines
\path(1350,2722)(1350,22)
\path(2250,2722)(2250,22)
\path(6450,2722)(6450,22)
\path(7350,2722)(7350,22)
\path(750,1822)(2550,1822)(3450,922)
\path(750,922)(2550,922)(3450,1822)

\path(5850,1822)(7650,1822)(8550,922)
\path(5850,922)(7650,922)(8550,1822)

\blacken\path(3393.198,1106.081)(3075,1297)(3265.919,978.802)(3393.198,1106.081)
\path(3075,1297)(3450,922)
\blacken\path(8493.198,1106.081)(8175,1297)(8365.919,978.802)(8493.198,1106.081)
\path(8175,1297)(8550,922)
\blacken\path(1440,2362)(1350,2722)(1260,2362)(1440,2362)
\path(1350,2722)(1350,2272)
\blacken\path(7440,2362)(7350,2722)(7260,2362)(7440,2362)
\path(7350,2722)(7350,2272)
\blacken\path(1260,382)(1350,22)(1440,382)(1260,382)
\path(1350,22)(1350,472)
\blacken\path(6360,382)(6450,22)(6540,382)(6360,382)
\path(6450,22)(6450,472)
\blacken\path(7260,382)(7350,22)(7440,382)(7260,382)
\path(7350,22)(7350,472)
\blacken\path(2160,382)(2250,22)(2340,382)(2160,382)
\path(2250,22)(2250,472)
\blacken\path(2340,2362)(2250,2722)(2160,2362)(2340,2362)
\path(2250,2722)(2250,2272)
\blacken\path(6540,2362)(6450,2722)(6360,2362)(6540,2362)
\path(6450,2722)(6450,2272)
\thinlines
\path(1650,2422)(1350,2122)
\path(1650,2422)(1350,2122)
\path(1950,2422)(1350,1822)
\path(1950,2422)(1350,1822)
\path(2250,2422)(1350,1522)
\path(2250,2422)(1350,1522)
\path(2250,2122)(1350,1222)
\path(2250,2122)(1350,1222)
\path(2250,1822)(1350,922)
\path(2250,1822)(1350,922)
\path(2250,1522)(1350,622)
\path(2250,1522)(1350,622)
\path(2250,1222)(1350,322)
\path(2250,1222)(1350,322)
\path(2250,922)(1650,322)
\path(2250,922)(1650,322)
\path(2250,622)(1950,322)
\path(2250,622)(1950,322)
\path(6750,2422)(6450,2122)
\path(6750,2422)(6450,2122)
\path(7050,2422)(6450,1822)
\path(7050,2422)(6450,1822)
\path(7350,2422)(6450,1522)
\path(7350,2422)(6450,1522)
\path(7350,2122)(6450,1222)
\path(7350,2122)(6450,1222)
\path(7350,1822)(6450,922)
\path(7350,1822)(6450,922)
\path(7350,1522)(6450,622)
\path(7350,1522)(6450,622)
\path(7350,1222)(6450,322)
\path(7350,1222)(6450,322)
\path(7350,922)(6750,322)
\path(7350,922)(6750,322)
\path(7350,622)(7050,322)
\path(7350,622)(7050,322)
\thicklines
\blacken\path(2868.198,1631.081)(2550,1822)(2740.919,1503.802)(2868.198,1631.081)
\path(2550,1822)(2925,1447)

\path(3000,1372)(2625,997)

\path(8100,1372)(7725,1747)
\blacken\path(7840.919,1240.198)(7650,922)(7968.198,1112.919)(7840.919,1240.198)
\path(7650,922)(8025,1297)
\path(750,1822)(1350,1822)
\blacken\path(990,1732)(1350,1822)(990,1912)(990,1732)
\path(750,922)(1350,922)
\blacken\path(990,832)(1350,922)(990,1012)(990,832)
\path(5850,922)(6450,922)
\blacken\path(6090,832)(6450,922)(6090,1012)(6090,832)
\path(5850,1822)(6450,1822)
\blacken\path(6090,1732)(6450,1822)(6090,1912)(6090,1732)

\put(4350,1372){$+      $}

\put(0000,1800){$x_{i-1}$}
\put(0000,0900){$x_{i  }$}

\put(3750,1800){$x_{i  }$}
\put(3750,0900){$x_{i-1}$}

\put(4950,1800){$x_{i-1}$}
\put(4950,0900){$x_{i  }$}
\put(8850,1800){$x_{i  }$}
\put(8850,0900){$x_{i-1}$}

\whiten\path(3259.081,1503.802)(3450,1822)(3131.802,1631.081)(3259.081,1503.802)
\whiten\path(8359.081,1503.802)(8550,1822)(8231.802,1631.081)(8359.081,1503.802)
\whiten\path(2809.081,1053.802)(3000,1372)(2681.802,1181.081)(2809.081,1053.802)
\whiten\path(7781.802,1562.919)(8100,1372)(7909.081,1690.198)(7781.802,1562.919)
\end{picture}

\begin{ca}
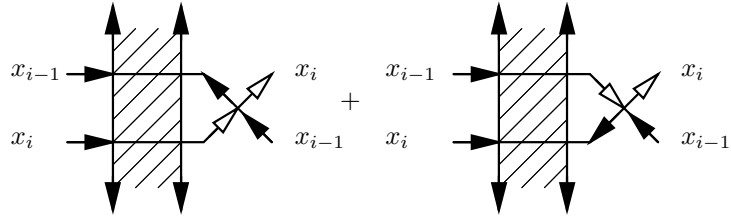
Multiplying the graph on the right by a type $b$ vertex and that 
on the left by a type $c$ vertex and adding the results.
\end{ca}

\end{minipage}
\end{center}

Since the arrows in the external loop on the right have both possible
orientations, they are summed over, and we have

\begin{center}
\begin{minipage}{4in}
\setlength{\unitlength}{0.001cm}

\begin{picture}(5000,3000)(-3000, 0)
\thicklines
\path(1350,2722)(1350,22)
\path(2250,2722)(2250,22)
\path(750,1822)(2550,1822)(3450,922)
\path(750,922)(2550,922)(3450,1822)
\whiten\path(3259.081,1503.802)(3450,1822)(3131.802,1631.081)(3259.081,1503.802)
\blacken\path(3393.198,1106.081)(3075,1297)(3265.919,978.802)(3393.198,1106.081)
\path(3075,1297)(3450,922)
\blacken\path(1440,2362)(1350,2722)(1260,2362)(1440,2362)
\path(1350,2722)(1350,2272)
\blacken\path(1260,382)(1350,22)(1440,382)(1260,382)
\path(1350,22)(1350,472)
\blacken\path(2160,382)(2250,22)(2340,382)(2160,382)
\path(2250,22)(2250,472)
\blacken\path(2340,2362)(2250,2722)(2160,2362)(2340,2362)
\path(2250,2722)(2250,2272)
\thinlines
\path(1650,2422)(1350,2122)
\path(1650,2422)(1350,2122)
\path(1950,2422)(1350,1822)
\path(1950,2422)(1350,1822)
\path(2250,2422)(1350,1522)
\path(2250,2422)(1350,1522)
\path(2250,2122)(1350,1222)
\path(2250,2122)(1350,1222)
\path(2250,1822)(1350,922)
\path(2250,1822)(1350,922)
\path(2250,1522)(1350,622)
\path(2250,1522)(1350,622)
\path(2250,1222)(1350,322)
\path(2250,1222)(1350,322)
\path(2250,922)(1650,322)
\path(2250,922)(1650,322)
\path(2250,622)(1950,322)
\path(2250,622)(1950,322)
\thicklines
\path(750,1822)(1350,1822)
\blacken\path(990,1732)(1350,1822)(990,1912)(990,1732)
\path(750,922)(1350,922)
\blacken\path(0990,0832)(1350,0922)(0990,1012)(0990,0832)

\put(0000,1822){$x_{i-1}$}
\put(0000,0922){$x_{i  }$}
\put(3750,1822){$x_{i  }$}
\put(3750,0922){$x_{i-1}$}

\end{picture}

\begin{ca}
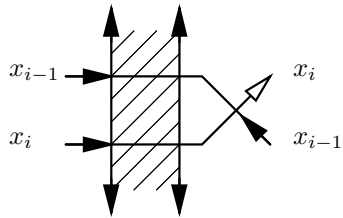
All possible arrow configurations that are allowed in the 
external loop on the right are summed over.
\end{ca}

\end{minipage}
\end{center}

But now we are in a position to use the Yang Baxter equation 
to move that external vertex, horizontally through the lattice, 
all the way to the left hand side


\begin{center}
\begin{minipage}{4in}
\setlength{\unitlength}{0.001cm}

\begin{picture}(3000,3000)(-3000, 0)
\thicklines
\blacken\path(1312,2362)(1222,2722)(1132,2362)(1312,2362)
\path(1222,2722)(1222,22)
\blacken\path(1132,382)(1222,22)(1312,382)(1132,382)
\blacken\path(2212,2362)(2122,2722)(2032,2362)(2212,2362)
\path(2122,2722)(2122,22)
\blacken\path(2032,382)(2122,22)(2212,382)(2032,382)
\path(2122,1822)(2722,1822)
\whiten\path(2362,1732)(2722,1822)(2362,1912)(2362,1732)
\blacken\path(2482,1012)(2122,922)(2482,832)(2482,1012)
\path(2122,922)(2722,922)
\thinlines
\path(1522,2422)(1222,2122)
\path(1822,2422)(1222,1822)
\path(2122,2422)(1222,1522)
\path(2122,2122)(1222,1222)
\path(2122,1822)(1222,922)
\path(2122,1522)(1222,622)
\path(2122,1222)(1222,322)
\path(2122,922)(1522,322)
\path(2122,622)(1822,322)
\thicklines
\path(1222,1822)(2122,1822)
\path(1222,922)(2122,922)
\path(1222,1822)(922,1822)
\path(1222,922)(922,922)
\path(922,922)(472,1372)
\path(922,1822)(472,1372)
\blacken\path(281.081,1053.802)(472,1372)(153.802,1181.081)(281.081,1053.802)
\path(472,1372)(22,922)
\blacken\path(153.802,1562.919)(472,1372)(281.081,1690.198)(153.802,1562.919)
\path(472,1372)(22,1822)

\put(-750,1800){$x_{i-1}$}
\put(-750,0900){$x_{i  }$}
\put(3000,1800){$x_{i  }$}
\put(3000,0900){$x_{i-1}$}

\end{picture}

\begin{ca}
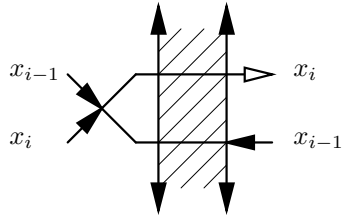
The result of using the Yang Baxter equation to weave the 
external loop all the way to the left. All possible arrow
configurations in that loop are summed over.
\end{ca}

\end{minipage}
\end{center}

Given the orientations of the boundary arrows on the left 
hand side, the vertex that emerges is uniquely a type 
$a$ vertex


\begin{center}
\begin{minipage}{4in}
\setlength{\unitlength}{0.001cm}

\begin{picture}(4500,3000)(-3000, 0)
\thicklines

\thicklines
\blacken\path(1890,2362)(1800,2722)(1710,2362)(1890,2362)
\path(1800,2722)(1800,22)
\blacken\path(1710,382)(1800,22)(1890,382)(1710,382)
\blacken\path(2790,2362)(2700,2722)(2610,2362)(2790,2362)
\path(2700,2722)(2700,22)
\blacken\path(2610,382)(2700,22)(2790,382)(2610,382)
\path(2700,1822)(3300,1822)
\whiten\path(2940,1732)(3300,1822)(2940,1912)(2940,1732)
\blacken\path(3060,1012)(2700,922)(3060,832)(3060,1012)
\path(2700,922)(3300,922)
\thinlines
\path(2100,2422)(1800,2122)
\path(2400,2422)(1800,1822)
\path(2700,2422)(1800,1522)
\path(2700,2122)(1800,1222)
\path(2700,1822)(1800,922)
\path(2700,1522)(1800,622)
\path(2700,1222)(1800,322)
\path(2700,922)(2100,322)
\path(2700,622)(2400,322)
\thicklines
\path(1800,1822)(2700,1822)
\path(1800,922)(2700,922)
\path(1500,1822)(1800,1822)
\path(1500,922)(1800,922)
\blacken\path(1309.081,1503.802)(1500,1822)(1181.802,1631.081)(1309.081,1503.802)
\path(1500,1822)(1050,1372)
\path(1050,1372)(1500,922)
\blacken\path(1181.802,1112.919)(1500,922)(1309.081,1240.198)(1181.802,1112.919)
\blacken\path(859.081,1053.802)(1050,1372)(731.802,1181.081)(859.081,1053.802)
\path(1050,1372)(600,922)
\blacken\path(731.802,1562.919)(1050,1372)(859.081,1690.198)(731.802,1562.919)
\path(1050,1372)(600,1822)

\put(-200,1800){$x_{i-1}$}
\put(-200,0900){$x_{i  }$}

\put(3600,1800){$x_{i  }$}
\put(3600,0900){$x_{i-1}$}

\end{picture}

\begin{ca}
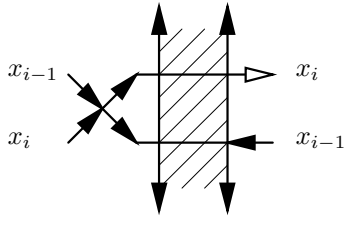
The only possible external vertex on the left is an $a$ vertex.
\end{ca}

\end{minipage}
\end{center}

Equating the initial sum with the final result, we obtain

\begin{equation}
\label{yetanotheridentityequation}
                     F[r_{i  }, x_{i  }] b(x_{i-1}, x_{i})
                   + F[r_{i-1}, x_{i-1}] c(x_{i-1}, x_{i})
= a(x_{i-1}, x_{i})  F[r_{i-1}, x_{i  }]
\end{equation}

Dividing both sides by $b(x_{i-1}, x_{i})$, moving the 
second term on the left hand side to the right, and using
$b(x_{i  }, x_{i-1}) =  - b(x_{i-1}, x_{i})$, and 
$c(x_{i  }, x_{i-1}) =    c(x_{i-1}, x_{i})$, we end
up with the identity

\begin{equation}
\label{mainequation}
F[r_{i}, x_{i}] = 
f(x_{i-1}, x_{i  })  F[r_{i-1}, x_{i  }] +
g(x_{i  }, x_{i-1})  F[r_{i-1}, x_{i-1}] 
\end{equation}
where we have defined
$$
f_{i, i+1} = \frac{a(x_{i}, x_{i+1})}{b(x_i, x_{i+1})}, \quad 
g_{i, i+1} = \frac{c(x_{i}, x_{i+1})}{b(x_i, x_{i+1})} 
$$

\begin{re}
We write the weight of a type $c$ vertex with formal 
dependence on its rapidities to make it easier to keep 
track of its position, although it is independent of 
the rapidities. 
\end{re}

Equation \ref{mainequation} says that we can rewrite 
the set of all configurations, in which the inverted 
arrow is on row $i$, in terms of configurations, 
in which the inverted arrow is in row ($i$$-$1). 

We have succeeded in `rolling' the inverted arrow by one
row towards the top. The price we have to pay for that is 
that we end up with two sets of configurations instead 
of one. 

Further, it easy to see that, if the initial inverted 
arrow is at row $r$, then rolling it upwards as 
we did above, the number of configurations will keep on 
doubling, leading to $2^{r-1}$ configurations.

However, as we will see below, the Yang Baxter equation
can be repeatedly used to reduce this number from $2^{r-1}$
to $r$, which is typical of how the Yang Baxter equation 
makes models solvable \cite{Baxter-book}.

To simplify the notation, we define 

\begin{equation}
\label{def1equation}
F_{i_1}^{i_2} = F[r_{i_1}, x_{i_2}]
\end{equation}
as the configuration with the arrow on the $i_1$-th horizontal 
line inverted, and the rapidity through that line is $x_2$, so 
we have

\begin{equation}
\label{basicequation}
F_{i}^{i} = f_{i-1, i  } F_{i-1}^{i  }
          + g_{i  , i-1} F_{i-1}^{i-1}
\end{equation}

\subsection{Rolling twice}

Suppose we roll an inverted arrow up twice. From equation 
\ref{basicequation}, we obtain

\begin{eqnarray*}
F_{i}^{i} &=& f_{i, i-1} F_{i-1}^{i  }
           +  g_{i-1, i} F_{i-1}^{i-1} \\
          &=& f_{  i, i-1} (   f_{  i, i-2} F_{i-2}^{i  } 
			     + g_{i-2, i  } F_{i-2}^{i-2} ) \\
          &+& g_{i-1,   i} (   f_{i-1, i-2} F_{i-2}^{i-1} 
	                     + g_{i-2, i-1} F_{i-2}^{i-2} ) 
\end{eqnarray*}
\begin{eqnarray*}
\phantom{F_{i}^{i}} &=& 
   f_{  i, i-1} f_{  i, i-2} F_{i-2}^{i  } 
+  f_{i-1, i-2} g_{i-1, i  } F_{i-2}^{i-1} 
\end{eqnarray*}
\begin{eqnarray}
\label{rollingtwiceequation}
\phantom{F_{i}^{i}} &+& 
  (g_{i-2,   i} f_{i,   i-1} + 
   g_{i-1,   i} g_{i-2, i-1}) F_{i-2}^{i-2}
\end{eqnarray}

Equation \ref{rollingtwiceequation} can be simplified by combining 
the two terms in the coefficient of $F_{i-2}^{i-2}$ into one using 
a Yang Baxter equation as follows. Re-expand the coefficients of 
$F_{i-2}^{i-2}$ in terms of Boltzmann weights to obtain

\begin{eqnarray}
\label{expandedequation}
g_{i-2,   i} f_{i, i-1} + g_{i-1, i} g_{i-2, i-1} &=& \\
\frac{c(x_{i-2}, x_{i  })}{b(x_{i-2}, x_{i  })} 
\frac{a(x_{i  }, x_{i-1})}{b(x_{i  }, x_{i-1})} 
&+& 
\frac{c(x_{i-1}, x_{i  })}{b(x_{i-1}, x_{i  })}
\frac{c(x_{i-2}, x_{i-1})}{b(x_{i-2}, x_{i-1})} 
\end{eqnarray}
Rewriting the right hand side of equation \ref{expandedequation}
in a slightly more convenient form, we obtain
\begin{eqnarray}
\label{twotermsequation}
\phantom{g_{i-2,   i} f_{i, i-1} + g_{i-1, i} g_{i-2, i-1}} 
&\phantom{=}& \\
-
\frac{c(x_{i  }, x_{i-2})}{b(x_{i  }, x_{i-2})}
\frac{a(x_{i  }, x_{i-1})}{b(x_{i  }, x_{i-1})}
&+&
\frac{c(x_{i  }, x_{i-1})}{b(x_{i  }, x_{i-1})}
\frac{c(x_{i-1}, x_{i-2})}{b(x_{i-1}, x_{i-2})}
\end{eqnarray}
Multiply both terms in equation \ref{twotermsequation} by 

\begin{equation}
\label{factorequation}
b(x_{i}, x_{i-1}) b(x_{i}, x_{i-2}) b(x_{i-1}, x_{i-2})
\end{equation}
we obtain
\begin{equation}
\label{resultingequation}
-
a(x_{i  }, x_{i-1}) c(x_{i  }, x_{i-2}) b(x_{i-1}, x_{i-2}) 
+
c(x_{i  }, x_{i-1}) c(x_{i-1}, x_{i-2}) b(x_{i  }, x_{i-2}) 
\end{equation}
Comparing equation \ref{resultingequation} with the left hand 
side of the Yang Baxter equation \ref{yangbaxterequation}, we 
find that they are identical, if we identify 
$x \rightarrow i, 
 y \rightarrow i-2, 
 z \rightarrow i-1$. Using these identifications, we can 
write equation \ref{yangbaxterequation} as

$$
c(x_{i-1}, x_{i-2}) c(x_{i  }, x_{i-1}) b(x_{i  }, x_{i-2})
+
b(x_{i-2}, x_{i-1}) a(x_{i  }, x_{i-1}) c(x_{i  }, x_{i-2}) 
$$
\begin{equation}
\label{newformybequation}
\phantom{c(x_{i-1}, x_{i-2}) c(x_{i  }, x_{i-1}) b(x_{i  }, x_{i-2})}
=
c(x_{i  }, x_{i-2}) b(x_{i  }, x_{i-1}) a(x_{i-2}, x_{i-1})
\end{equation}

Dividing the right hand side of equation \ref{newformybequation} 
by the expression in equation \ref{factorequation}, we obtain

$$
g_{i-2, i} f_{i,   i-1} + g_{i-2, i-1} g_{i-1, i} =
g_{i-2, i} f_{i-2, i-1} 
$$
So we end up with
\begin{equation}
\label{resultofrollingtwiceequation}
F_{i}^{i} =
   f_{  i, i-1} f_{  i, i-2} F_{i-2}^{i  }
 + f_{i-1, i-2} g_{i-1, i  } F_{i-2}^{i-1}
 + f_{i-2, i-1} g_{i-2, i  } F_{i-2}^{i-2}
\end{equation}

\subsection{Rolling many times}

It is straightforward to iterate the above equation to bring 
the inverted arrow to the top. If the initial position is $r$, 
with rapidity $x_r$, we obtain

$$
F_{r}^{r} = 
\sum_{\alpha=1}^{r-1} 
\ll 
g_{\alpha, r} \prod_{i = 1 \atop {i \ne \alpha}}^{r-1} f_{\alpha, i} 
\rr
F_{1}^{\alpha} 
+ 
\ll
\prod_{i=1}^{r-1} f_{r, i}
\rr
F_{1}^{r}
= 
\sum_{\alpha=1}^r 
\ll
\frac{g_{\alpha, r}}{f_{\alpha, r}}
\prod_{i = 1 \atop {i \ne \alpha}}^{r} f_{\alpha, i} 
\rr
F_{1}^{\alpha} 
$$
where we have used
$$
\frac{g_{r,  r}}{f_{r,  r}} = 1
$$

Initially, we have the rows all labelled correctly, but the arrow 
on row $r$ is inverted. 

After rolling the inverted arrow (in the remaining lattice) to the 
top row, as we did above, the top row (which has $n-1$ vertices) 
is now `frozen', in the sense that we know that all of its vertices 
are $a$ vertices. Peeling these, we pick up a factor of 
$$
\prod_{j=2}^{N} a(x_k, y_j)
$$
where $x_k$ is the rapidity that ended at the top.
The partition function of the remaining configurations can now be 
computed using Izergin's expression. Putting all contributions 
together, we obtain
$$
F_{r}^{r} = 
\sum_{\alpha=1}^r
\ll 
\prod_{j=2}^{N} a(x_{\alpha}, y_{j}) 
\rr
\ll
\frac{g_{\alpha, r}}{f_{\alpha, r}}
\prod_{i=1 \atop {i \ne \alpha}}^{r} f_{\alpha, i}
\rr
Z_{N-1}\ll \underline{x}_{\alpha}, \underline{y}_{1}\rr
$$
where $Z_{N-1}\ll \underline{x}_{\alpha}, \underline{y}_{1}\rr$,
with underlined arguments, is Izergin's partition function for 
an ($N$$-$$1$)$\times$($N$$-$$1$) lattice, with the same 
assignment of horizontal and vertical rapidities, as those 
of $Z_{N}\ll \{x\}, \{y\}\rr $, but with $x_{\alpha}$ and $y_{1}$ 
missing. 

\paragraph*{The result of Bogoliubov, Pronko and Zvonarev} for 
the boundary 1-point function, $H_{N}^{r}$, in their definition,
can be produced by multiplying the left and the right 1-point 
functions computed above, and normalising the product by the 
partition function of the $N$$\times$$N$ model, to obtain

$$
H_{N}^{r}\ll \{x\}, \{y\} \rr = 
\frac{1}{Z_N\ll \{x\}, \{y\} \rr}
\ll \prod_{i=1  }^{r-1} b(x_r, y_1) \rr
                                 c(x_r,y_1)
\ll \prod_{i=r+1}^{N  } a(x_i,y_1) \rr \times 
$$
\begin{equation}
\label{bpzequation}
\sum_{\alpha=1}^r
\ll
\prod_{j=2}^{N} a(x_{\alpha}, y_{j})
\rr
\ll
\frac{g_{\alpha, r}}{f_{\alpha, r}}
\prod_{i=1 \atop {i \ne \alpha}}^{r} f_{\alpha, i}
\rr
Z_{N-1}\ll \underline{x}_{\alpha}, \underline{y}_{1}\rr
\end{equation}

\subsection{Rolling down}

It is possible to repeat the above exercise by rolling downwards,
rather than upwards. The resulting expression will look different 
from that obtained above. In particular, the peeled rows will be 
products of $b$ rather than $a$ weights. We leave it as an (easy) 
exercise to the reader to show that the resulting expression is 
trivially equivalent to the first.   

\section{Simple boundary 2-point functions}

We are now ready to extend the above arguments to simple boundary 
$2$-point functions. There are four cases to consider: 

\begin{enumerate}
\item The second inverted arrow is on the same boundary as 
      the first.  
\item The second inverted arrow is on one of the two boundaries 
      adjacent (rather than opposite) to that of the first. 
\item The second inverted arrow is on the opposite boundary 
      to that of the first, but not directly opposite to it. 
\item The second inverted arrow is on the opposite boundary 
      to that of the first, and directly opposite to it. 
\end{enumerate}

Case 1 arises when an $N$$\times$$N$ domain wall lattice is 
sliced into two parts with one part having $2$$\times$$N$ 
vertical (or horizontal lines) while the other part has the 
rest.

Cases 2, 3 and 4 arise in more complicated dissections of 
the original domain wall lattice (for example, into more 
than 2 parts). We deal with them for completeness.

\subsection{Case 1}

A non-trivial example of a simple 2-point function, with both 
inverted arrows on the same side, can be drawn schematically 
as follows


\begin{center}
\begin{minipage}{2.5in}
\setlength{\unitlength}{0.001cm}

\begin{picture}(4000,6000)(-1000, 0)
\thicklines

\path(1200,5700)(1200,300)
\path(2100,5700)(2100,300)
\path(3000,5700)(3000,300)
\path(300,4800)(3900,4800)
\path(300,3000)(3900,3000)
\path(300,2100)(3900,2100)
\path(300,1200)(3900,1200)
\path(600,4800)(1200,4800)
\blacken\path(840,4710)(1200,4800)(840,4890)(840,4710)
\path(600,3900)(1200,3900)
\blacken\path(840,3810)(1200,3900)(840,3990)(840,3810)
\path(600,3000)(1200,3000)
\blacken\path(840,2910)(1200,3000)(840,3090)(840,2910)
\path(600,2100)(1200,2100)
\blacken\path(840,2010)(1200,2100)(840,2190)(840,2010)
\path(600,1200)(1200,1200)
\blacken\path(840,1110)(1200,1200)(840,1290)(840,1110)
\path(3600,4800)(3000,4800)
\blacken\path(3360,4890)(3000,4800)(3360,4710)(3360,4890)
\path(3600,1200)(3000,1200)
\blacken\path(3360,1290)(3000,1200)(3360,1110)(3360,1290)
\path(1200,5100)(1200,5700)
\blacken\path(1290,5340)(1200,5700)(1110,5340)(1290,5340)
\path(2100,5100)(2100,5700)
\blacken\path(2190,5340)(2100,5700)(2010,5340)(2190,5340)
\path(3000,5100)(3000,5700)
\blacken\path(3090,5340)(3000,5700)(2910,5340)(3090,5340)
\path(1200,825)(1200,225)
\blacken\path(1110,585)(1200,225)(1290,585)(1110,585)
\path(2100,825)(2100,225)
\blacken\path(2010,585)(2100,225)(2190,585)(2010,585)
\path(3000,825)(3000,225)
\blacken\path(2910,585)(3000,225)(3090,585)(2910,585)
\path(3300,2100)(3900,2100)
\whiten\path(3540,2010)(3900,2100)(3540,2190)(3540,2010)
\path(300,3900)(3900,3900)
\path(3300,3900)(3900,3900)
\whiten\path(3540,3810)(3900,3900)(3540,3990)(3540,3810)
\path(3600,3000)(3000,3000)
\blacken\path(3360,3090)(3000,3000)(3360,2910)(3360,3090)

\put(-100,4800){$1  $}
\put(-100,1200){$N  $}

\put(4100,3900){$r_1$}
\put(4100,2100){$r_1$}

\put(2900,-100){$3$}
\put(1100,-100){$N$}

\end{picture}

\begin{ca}
A simple boundary 2-point function.
\end{ca}

\end{minipage}
\end{center}

If the initial positions of the inverted arrows are $r_1$ and $r_2$,
where $r_1 < r_2$, with rapidities $x_{r_1}$ and $x_{r_2}$,
respectively, we can roll the upper arrow to the top, then roll the 
lower arrow to the row below the first, and obtain

$$
F_{r_1, r_2}^{r_1, r_2} = 
\sum_{\alpha_{1} = 1}^{r_1} 
\ll 
\frac{g_{\alpha_{1}, r_1}}{f_{\alpha_{1}, r_1}} 
\prod_{i_1=1 \atop {i_1 \ne \alpha_{1}}}^{r_1} f_{\alpha_{1}, i_1} 
\rr 
\ll \prod_{j=3}^{N} a(x_{\alpha_{1}}, y_j) \rr
F_{r_2}^{r_2}
$$
and
$$
F_{r_2}^{r_2} =
\sum_{\alpha_{2}=2 \atop{\alpha_{2} \ne \alpha_{1}}}^{r_2} 
\ll \frac{g_{\alpha_{2}, r_2}}{f_{\alpha_{2}, r_2}}
    \prod_{i_2=2 \atop{i_2 \ne \alpha_{2}}}^{r_2} f_{\alpha_{2}, i_2} 
\rr
\ll 
\prod_{j=3}^{N} a(x_{\alpha_{2}}, y_j) 
\rr
Z_{N-2}\ll \underline{x}_{\alpha_{1}}, 
           \underline{x}_{\alpha_{2}},
           \underline{y}_{1         }, 
	   \underline{y}_{2         }\rr
$$
The full result is obtained by composing the above expressions.
\phantom{{}}The extension of the above to ($N$$-$$n$)$\times$$N$ 
lattices with $n$ inverted arrows, all on the same boundary, is 
clear.

\subsection{Case 2}

In this case, there are two types of configurations to consider. 
In both types, we are dealing initially with an $N$$\times$$N$ 
lattice. The first type can be drawn schematically as 


\begin{center}
\begin{minipage}{4in}
\setlength{\unitlength}{0.001cm}
\begin{picture}(6000, 6000)(-2000, 0)

\thicklines

\path(525,3954)(1125,3954)

\blacken\path(765,3864)(1125,3954)(765,4044)(765,3864)
\path(600,3054)(1200,3054)
\blacken\path(840,2964)(1200,3054)(840,3144)(840,2964)
\path(600,2154)(1200,2154)
\blacken\path(840,2064)(1200,2154)(840,2244)(840,2064)
\path(600,1254)(1200,1254)
\blacken\path(840,1164)(1200,1254)(840,1344)(840,1164)
\path(1200,1029)(1200,429)
\blacken\path(1110,789)(1200,429)(1290,789)(1110,789)
\path(2100,954)(2100,354)
\blacken\path(2010,714)(2100,354)(2190,714)(2010,714)
\path(3000,954)(3000,354)
\blacken\path(2910,714)(3000,354)(3090,714)(2910,714)
\path(3900,954)(3900,354)
\blacken\path(3810,714)(3900,354)(3990,714)(3810,714)
\path(300,3954)(4800,3954)
\path(300,3054)(4800,3054)
\path(300,2154)(4800,2154)
\path(300,1254)(4800,1254)
\path(4500,3054)(3900,3054)
\blacken\path(4260,3144)(3900,3054)(4260,2964)(4260,3144)
\path(4425,1254)(3825,1254)
\blacken\path(4185,1344)(3825,1254)(4185,1164)(4185,1344)
\path(1200,4854)(1200,804)
\path(2100,4854)(2100,804)
\path(3000,4854)(3000,804)
\path(1200,4254)(1200,4854)
\blacken\path(1290,4494)(1200,4854)(1110,4494)(1290,4494)
\path(2100,4254)(2100,4854)
\blacken\path(2190,4494)(2100,4854)(2010,4494)(2190,4494)
\path(3900,4254)(3900,4854)
\blacken\path(3990,4494)(3900,4854)(3810,4494)(3990,4494)
\path(3000,4554)(3000,3954)
\whiten\path(2910,4314)(3000,3954)(3090,4314)(2910,4314)
\path(3900,4854)(3900,804)
\path(4200,2154)(4800,2154)
\whiten\path(4440,2064)(4800,2154)(4440,2244)(4440,2064)
\path(4500,3954)(3900,3954)
\blacken\path(4260,4044)(3900,3954)(4260,3864)(4260,4044)

\put(0000,3950){$2$}
\put(0000,3050){$3$}
\put(0000,1250){$N$}

\put(1100,-050){$N$}
\put(2900,-050){$c$}
\put(3800,-050){$2$}

\put(1150,5100){$y_N$}
\put(2950,5100){$y_c$}
\put(3850,5100){$y_2$}

\put(5100,3950){$x_2$}
\put(5100,2150){$x_r$}
\put(5100,1250){$x_N$}

\end{picture}

\begin{ca}
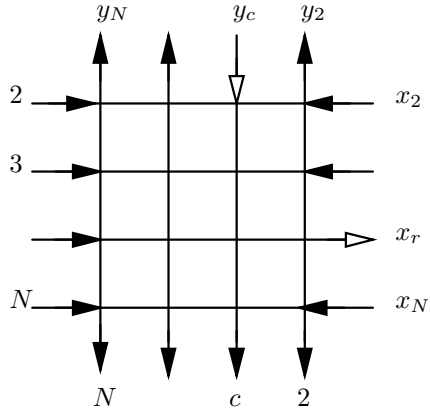
A simple boundary 2-point function with inverted arrows 
on orthogonal boundaries.
\end{ca}

\end{minipage}
\end{center}

In the second type, we consider a horizontal inverted arrow pointing 
to the right boundary, and a vertical inverted arrow pointing 
away from the lower boundary. Both types can be treated analogously 
(leading to analogous results with minor differences).  
In the following we will deal with the first and leave the second
as an exercise to the reader.

If the positions of the inverted arrows are row $r$ and column $c$, 
then we can write the result in two steps. First, we roll the 
inverted arrow on the right boundary all the way to the top to 
obtain\footnote{In this case there is no simple frozen 
upper row that can be peeled.}

$$
F_{c r}^{c r} = 
\sum_{\alpha = 2}^{r}
\ll
\frac{g_{\alpha, r}}{f_{\alpha, r}}
\prod^{r}_{i = 2 \atop{i \ne \alpha}} f_{\alpha, i}
\rr
F_{c 1}^{c \alpha}
$$
This leaves us with the factor $F_{c \alpha}^{c 1}$ corresponding 
to the lattice 


\begin{center}
\begin{minipage}{4in}

\setlength{\unitlength}{0.001cm}
\begin{picture}(6000, 6000)(-2000, 0)

\thicklines
\path(525,3954)(1125,3954)
\blacken\path(765,3864)(1125,3954)(765,4044)(765,3864)
\path(600,3054)(1200,3054)
\blacken\path(840,2964)(1200,3054)(840,3144)(840,2964)
\path(600,2154)(1200,2154)
\blacken\path(840,2064)(1200,2154)(840,2244)(840,2064)
\path(600,1254)(1200,1254)
\blacken\path(840,1164)(1200,1254)(840,1344)(840,1164)
\path(1200,1029)(1200,429)
\blacken\path(1110,789)(1200,429)(1290,789)(1110,789)
\path(2100,954)(2100,354)
\blacken\path(2010,714)(2100,354)(2190,714)(2010,714)
\path(3000,954)(3000,354)
\blacken\path(2910,714)(3000,354)(3090,714)(2910,714)
\path(3900,954)(3900,354)
\blacken\path(3810,714)(3900,354)(3990,714)(3810,714)
\path(300,3954)(4800,3954)
\path(300,3054)(4800,3054)
\path(300,2154)(4800,2154)
\path(300,1254)(4800,1254)
\path(4500,3054)(3900,3054)
\blacken\path(4260,3144)(3900,3054)(4260,2964)(4260,3144)
\path(4425,1254)(3825,1254)
\blacken\path(4185,1344)(3825,1254)(4185,1164)(4185,1344)
\path(1200,4854)(1200,804)
\path(2100,4854)(2100,804)
\path(3000,4854)(3000,804)
\path(2100,4254)(2100,4854)
\blacken\path(2190,4494)(2100,4854)(2010,4494)(2190,4494)
\path(3900,4254)(3900,4854)
\blacken\path(3990,4494)(3900,4854)(3810,4494)(3990,4494)
\path(3900,4854)(3900,804)
\path(4500,2154)(3900,2154)
\blacken\path(4260,2244)(3900,2154)(4260,2064)(4260,2244)
\path(4200,3954)(4800,3954)
\whiten\path(4440,3864)(4800,3954)(4440,4044)(4440,3864)
\path(1200,4254)(1200,4854)
\blacken\path(1290,4494)(1200,4854)(1110,4494)(1290,4494)
\path(3000,4554)(3000,3954)

\whiten\path(2910,4314)(3000,3954)(3090,4314)(2910,4314)

\put(0000,3950){$2$}
\put(5000,3950){$x_{\alpha}$}
\put(0000,3050){$3$}
\put(0000,1250){$N$}

\put(1100,0000){$N$}
\put(2900,0000){$c$}
\put(3800,0000){$2$}

\put(1200,5150){$y_N$}
\put(3000,5150){$y_c$}
\put(3900,5150){$y_2$}

\end{picture}

\begin{ca}
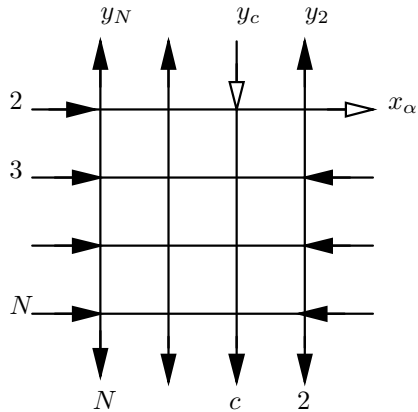
The right inverted arrow has been rolled all the way to the top.
\end{ca}

\end{minipage}
\end{center}

We need to roll the vertical inverted arrow all the way to the 
left. This can be done using exactly the same method that was 
used above to roll a horizontal inverted arrow to the top. In 
this case, we need to `inject' an $a$ vertex into the lattice
from below and use the Yang Baxter equation to thread it all
the way to the top. The result is exactly the same as in the 
case of horizontal arrows given that the extra vertex will move 
through the lattice in the same direction as the orientations 
of {\it both} vertical rapidities that end up swapping positions. 
The resulting configurations


\begin{center}
\begin{minipage}{4in}
\setlength{\unitlength}{0.001cm}

\begin{picture}(6000, 6000)(-2000, 0)
\thicklines
\path(525,3954)(1125,3954)
\blacken\path(765,3864)(1125,3954)(765,4044)(765,3864)
\path(600,3054)(1200,3054)
\blacken\path(840,2964)(1200,3054)(840,3144)(840,2964)
\path(600,2154)(1200,2154)
\blacken\path(840,2064)(1200,2154)(840,2244)(840,2064)
\path(600,1254)(1200,1254)
\blacken\path(840,1164)(1200,1254)(840,1344)(840,1164)
\path(1200,1029)(1200,429)
\blacken\path(1110,789)(1200,429)(1290,789)(1110,789)
\path(2100,954)(2100,354)
\blacken\path(2010,714)(2100,354)(2190,714)(2010,714)
\path(3000,954)(3000,354)
\blacken\path(2910,714)(3000,354)(3090,714)(2910,714)
\path(3900,954)(3900,354)
\blacken\path(3810,714)(3900,354)(3990,714)(3810,714)
\path(300,3954)(4800,3954)
\path(300,3054)(4800,3054)
\path(300,2154)(4800,2154)
\path(300,1254)(4800,1254)
\path(4500,3054)(3900,3054)
\blacken\path(4260,3144)(3900,3054)(4260,2964)(4260,3144)
\path(4425,1254)(3825,1254)
\blacken\path(4185,1344)(3825,1254)(4185,1164)(4185,1344)
\path(1200,4854)(1200,804)
\path(2100,4854)(2100,804)
\path(3000,4854)(3000,804)
\path(2100,4254)(2100,4854)
\blacken\path(2190,4494)(2100,4854)(2010,4494)(2190,4494)
\path(3900,4254)(3900,4854)
\blacken\path(3990,4494)(3900,4854)(3810,4494)(3990,4494)
\path(3900,4854)(3900,804)
\path(4500,2154)(3900,2154)
\blacken\path(4260,2244)(3900,2154)(4260,2064)(4260,2244)
\path(4200,3954)(4800,3954)
\whiten\path(4440,3864)(4800,3954)(4440,4044)(4440,3864)
\path(1200,4554)(1200,3954)
\whiten\path(1110,4314)(1200,3954)(1290,4314)(1110,4314)
\path(3000,4254)(3000,4854)
\blacken\path(3090,4494)(3000,4854)(2910,4494)(3090,4494)

\put(-100,3950){$2$}
\put(-100,3050){$3$}
\put(-100,1250){$N$}

\put(1100,5100){$y_\beta$}

\put(3800,5100){$y_2$}

\put(1100,0000){$N$}
\put(2900,0000){$c$}
\put(3800,0000){$2$}

\put(5000,3950){$x_{\alpha}$}

\end{picture}

\begin{ca}
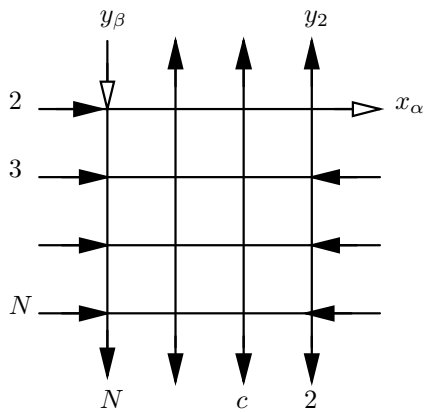
The upper inverted arrow has been rolled all the way to the left.
\end{ca}

\end{minipage}
\end{center}
can be evaluated by inspection to be

\begin{eqnarray}
\label{someequation}
F_{c 1}^{c \alpha}=
\sum_{\beta=c}^{N} 
\ll 
\frac{\tilde{g}_{\beta, c}}{\tilde{f}_{\beta, c}} 
\prod_{j=c \atop{j \ne \beta}}^{N  } \tilde{f}_{\beta,  j}
\rr
\ll
\prod_{j=2 \atop{j \ne \beta}}^{N-1} a(x_\alpha, y_j)
\rr
\ll
\prod_{i=2}^{N} b(x_i, y_\beta)
\rr \times \\
Z_{N-2}\ll \underline{x}_{1     }, 
           \underline{y}_{1     },
           \underline{x}_{\alpha}, 
	   \underline{y}_{\beta }\rr
\end{eqnarray}
where we have defined
$$
\tilde{f}_{i, i+1} = \frac{a(y_{i}, y_{i+1})}{b(y_i, y_{i+1})}, \quad 
\tilde{g}_{i, i+1} = \frac{c(y_{i}, y_{i+1})}{b(y_i, y_{i+1})} 
$$
\subsection{Case 3}

This is a direct extension of case 1. We take the inverted 
arrows to be those on the right and left boundaries. 


\begin{center}
\begin{minipage}{4in}
\setlength{\unitlength}{0.001cm}

\begin{picture}(6000,6000)(-2000, 0)
\thicklines
\path(1200,5700)(1200,300)
\path(2100,5700)(2100,300)
\path(3000,5700)(3000,300)
\path(3900,5700)(3900,300)
\path(300,4800)(4800,4800)
\path(300,3900)(4800,3900)
\path(300,3000)(4800,3000)
\path(300,2100)(4800,2100)
\path(300,1200)(4800,1200)
\path(1200,5100)(1200,5700)
\blacken\path(1290,5340)(1200,5700)(1110,5340)(1290,5340)
\path(2100,5100)(2100,5700)
\blacken\path(2190,5340)(2100,5700)(2010,5340)(2190,5340)
\path(3000,5100)(3000,5700)
\blacken\path(3090,5340)(3000,5700)(2910,5340)(3090,5340)
\path(3900,5100)(3900,5700)
\blacken\path(3990,5340)(3900,5700)(3810,5340)(3990,5340)
\path(1200,900)(1200,300)
\blacken\path(1110,660)(1200,300)(1290,660)(1110,660)
\path(2100,900)(2100,300)
\blacken\path(2010,660)(2100,300)(2190,660)(2010,660)
\path(3000,900)(3000,300)
\blacken\path(2910,660)(3000,300)(3090,660)(2910,660)
\path(3900,900)(3900,300)
\blacken\path(3810,660)(3900,300)(3990,660)(3810,660)
\path(4425,4800)(3825,4800)
\blacken\path(4185,4890)(3825,4800)(4185,4710)(4185,4890)
\path(4425,3900)(3825,3900)
\blacken\path(4185,3990)(3825,3900)(4185,3810)(4185,3990)
\path(4425,2100)(3825,2100)
\blacken\path(4185,2190)(3825,2100)(4185,2010)(4185,2190)
\path(4425,1200)(3825,1200)
\blacken\path(4185,1290)(3825,1200)(4185,1110)(4185,1290)
\path(600,4800)(1200,4800)
\blacken\path(840,4710)(1200,4800)(840,4890)(840,4710)
\path(600,3900)(1200,3900)
\blacken\path(840,3810)(1200,3900)(840,3990)(840,3810)
\path(600,3000)(1200,3000)
\blacken\path(840,2910)(1200,3000)(840,3090)(840,2910)
\path(600,1200)(1200,1200)
\blacken\path(840,1110)(1200,1200)(840,1290)(840,1110)
\path(4200,3000)(4800,3000)
\whiten\path(4440,2910)(4800,3000)(4440,3090)(4440,2910)
\path(900,2100)(300,2100)
\whiten\path(660,2190)(300,2100)(660,2010)(660,2190)

\put(5100,4800){$1  $}
\put(5100,3000){$r_1$}
\put(-200,2100){$r_2$}
\put(5100,1200){$N  $}

\put(1100,-100){$N  $}
\put(2900,-100){$4  $}
\put(3800,-100){$3  $}

\end{picture}

\begin{ca}
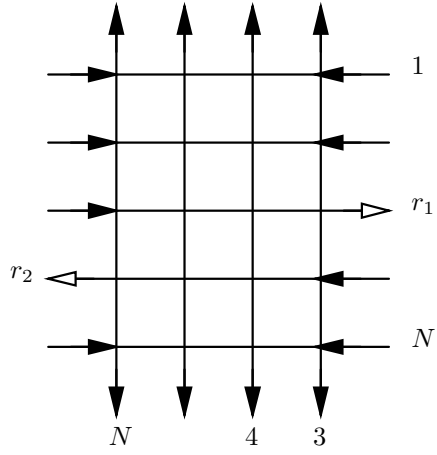
A simple boundary 2-point function with inverted arrows 
on opposite boundaries, but not directly opposite to one 
another. 
\end{ca}

\end{minipage}
\end{center}

Rolling the higher inverted arrow, then the lower one, we end 
up with the configuration 


\begin{center}
\begin{minipage}{4in}
\setlength{\unitlength}{0.001cm}

\begin{picture}(5000, 6000)(-2000, 0)
\thicklines
\path(997,5422)(997,22)
\path(1897,5422)(1897,22)
\path(2797,5422)(2797,22)
\path(3697,5422)(3697,22)
\path(97,4522)(4597,4522)
\path(97,3622)(4597,3622)
\path(97,2722)(4597,2722)
\path(97,1822)(4597,1822)
\path(97,922)(4597,922)
\path(997,4822)(997,5422)
\blacken\path(1087,5062)(997,5422)(907,5062)(1087,5062)
\path(1897,4822)(1897,5422)
\blacken\path(1987,5062)(1897,5422)(1807,5062)(1987,5062)
\path(2797,4822)(2797,5422)
\blacken\path(2887,5062)(2797,5422)(2707,5062)(2887,5062)
\path(3697,4822)(3697,5422)
\blacken\path(3787,5062)(3697,5422)(3607,5062)(3787,5062)
\path(997,622)(997,22)
\blacken\path(907,382)(997,22)(1087,382)(907,382)
\path(1897,622)(1897,22)
\blacken\path(1807,382)(1897,22)(1987,382)(1807,382)
\path(2797,622)(2797,22)
\blacken\path(2707,382)(2797,22)(2887,382)(2707,382)
\path(3697,622)(3697,22)
\blacken\path(3607,382)(3697,22)(3787,382)(3607,382)
\path(4222,3622)(3622,3622)
\blacken\path(3982,3712)(3622,3622)(3982,3532)(3982,3712)
\path(4222,1822)(3622,1822)
\blacken\path(3982,1912)(3622,1822)(3982,1732)(3982,1912)
\path(4222,922)(3622,922)
\blacken\path(3982,1012)(3622,922)(3982,832)(3982,1012)
\path(397,4522)(997,4522)
\blacken\path(637,4432)(997,4522)(637,4612)(637,4432)
\path(397,2722)(997,2722)
\blacken\path(637,2632)(997,2722)(637,2812)(637,2632)
\path(397,922)(997,922)
\blacken\path(637,832)(997,922)(637,1012)(637,832)
\path(622,3622)(22,3622)
\whiten\path(382,3712)(22,3622)(382,3532)(382,3712)
\path(3997,4522)(4597,4522)
\whiten\path(4237,4432)(4597,4522)(4237,4612)(4237,4432)
\path(4297,2722)(3697,2722)
\blacken\path(4057,2812)(3697,2722)(4057,2632)(4057,2812)
\path(397,1822)(997,1822)
\blacken\path(637,1732)(997,1822)(637,1912)(637,1732)

\put(4800,4500){$x_{\alpha_1}$}
\put(-800,3600){$x_{\alpha_2}$}

\end{picture}

\begin{ca}
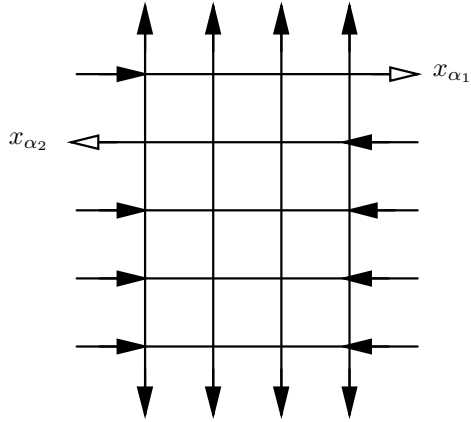
The inverted arrow have been rolled all the way to the 
top two rows.
\end{ca}

\end{minipage}
\end{center}
which can be evaluated by inspection to be

$$
\ll {}_{r_2}^{r_2}F_{r_1}^{r_1} \rr =
\sum_{\alpha_1 = 1}^{r_1}
\ll
\frac{g_{\alpha_1, r_1}}{f_{\alpha_1, r_1}}
\prod_{i=1 \atop {i \ne \alpha_1}}^{r_1} f_{\alpha_1, i}
\rr
\ll \prod_{j=3}^{N} a(x_{\alpha_1}, y_j) \rr
\ll {}_{r_2}^{r_2}F \rr
$$
where
$$
\ll {}_{r_2}^{r_2}F \rr =
\sum_{\alpha_2=2 \atop{\alpha_2 \ne \alpha_1}}^{r_2}
\ll \frac{g_{\alpha_2, r_2}}{f_{\alpha_2, r_2}}
\prod_{i_2 = 2 \atop{j \ne r_1}}^{r_2} f_{\alpha_2, i_2}
\rr
\ll 
\prod_{j=3}^{N} b(x_{\alpha_2}, y_j) 
\rr
Z_{N-2}\ll \underline{x}_{\alpha_1}, 
           \underline{x}_{\alpha_2},
           \underline{y}_{1       }, 
	   \underline{y}_{2       }\rr
$$
where we have used a self-explanatory notation to indicate
the position and rapidity labels of the inverted arrow on 
the left boundary.

\subsection{Case 4}

In this case, we need to deal with inverted arrows that 
are directly opposite to each other.


\begin{center}
\begin{minipage}{4in}
\setlength{\unitlength}{0.001cm}

\begin{picture}(5000,6000)(-2000, 0)
\thicklines
\path(0922,5422)(0922,0022)
\path(1822,5422)(1822,0022)
\path(2722,5422)(2722,0022)
\path(3622,5422)(3622,0022)
\path(0022,4522)(4522,4522)
\path(0022,3622)(4522,3622)
\path(0022,2722)(4522,2722)
\path(0022,1822)(4522,1822)
\path(0022,0922)(4522,0922)
\path(0922,4822)(0922,5422)
\blacken\path(1012,5062)(922,5422)(832,5062)(1012,5062)
\path(1822,4822)(1822,5422)
\blacken\path(1912,5062)(1822,5422)(1732,5062)(1912,5062)
\path(2722,4822)(2722,5422)
\blacken\path(2812,5062)(2722,5422)(2632,5062)(2812,5062)
\path(3622,4822)(3622,5422)
\blacken\path(3712,5062)(3622,5422)(3532,5062)(3712,5062)
\path(922,622)(922,22)
\blacken\path(832,382)(922,22)(1012,382)(832,382)
\path(1822,622)(1822,22)
\blacken\path(1732,382)(1822,22)(1912,382)(1732,382)
\path(2722,622)(2722,22)
\blacken\path(2632,382)(2722,22)(2812,382)(2632,382)
\path(3622,622)(3622,22)
\blacken\path(3532,382)(3622,22)(3712,382)(3532,382)
\path(4147,3622)(3547,3622)
\blacken\path(3907,3712)(3547,3622)(3907,3532)(3907,3712)
\path(4147,1822)(3547,1822)
\blacken\path(3907,1912)(3547,1822)(3907,1732)(3907,1912)
\path(4147,922)(3547,922)
\blacken\path(3907,1012)(3547,922)(3907,832)(3907,1012)
\path(322,4522)(922,4522)
\blacken\path(562,4432)(922,4522)(562,4612)(562,4432)
\path(322,922)(922,922)
\blacken\path(562,832)(922,922)(562,1012)(562,832)
\path(322,1822)(922,1822)
\blacken\path(562,1732)(922,1822)(562,1912)(562,1732)
\path(322,3622)(922,3622)
\blacken\path(562,3532)(922,3622)(562,3712)(562,3532)
\path(4222,4522)(3622,4522)
\blacken\path(3982,4612)(3622,4522)(3982,4432)(3982,4612)
\path(622,2722)(22,2722)
\path(3922,2722)(4522,2722)

\whiten\path(382,2812)(22,2722)(382,2632)(382,2812)
\whiten\path(4162,2632)(4522,2722)(4162,2812)(4162,2632)

\put(-500,2700){$x_i$}
\put(4800,2700){$  i$}

\end{picture}

\begin{ca}
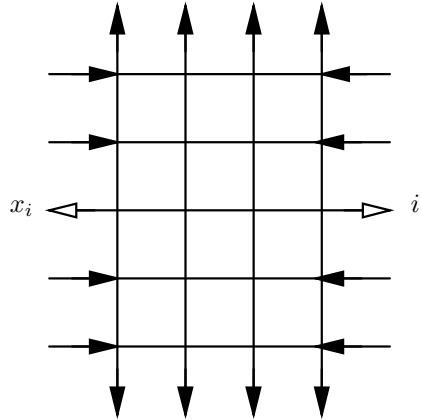
A 2-point function with two inverted arrows on opposite 
boundaries, and directly opposite to one another.
\end{ca}

\end{minipage}
\end{center}

It is clear that in this case, the above rolling argument needs 
to be slightly modified, as there in no way to multiply by an $a$ 
weight from one side, and thread that to the other side to roll 
an inverted arrow.

We wish to show that this case can be reduced to a sum over cases
of type 3, discussed above. Consider two configurations 
drawn schematically as follows (all external arrows that are not
drawn are meant to be the same in both configurations, and all 
internal arrows are summed over)


\begin{center}
\begin{minipage}{4in}
\setlength{\unitlength}{0.001cm}

\begin{picture}(8000,3000)(-2000, 0)
\thicklines

\path(1050,2722)(1050,22)
\path(1950,2722)(1950,22)
\path(6150,2722)(6150,22)
\path(7050,2722)(7050,22)
\blacken\path(1140,2362)(1050,2722)(960,2362)(1140,2362)
\path(1050,2722)(1050,2272)
\blacken\path(7140,2362)(7050,2722)(6960,2362)(7140,2362)
\path(7050,2722)(7050,2272)
\blacken\path(960,382)(1050,22)(1140,382)(960,382)
\path(1050,22)(1050,472)
\blacken\path(6060,382)(6150,22)(6240,382)(6060,382)
\path(6150,22)(6150,472)
\blacken\path(6960,382)(7050,22)(7140,382)(6960,382)
\path(7050,22)(7050,472)
\blacken\path(1860,382)(1950,22)(2040,382)(1860,382)
\path(1950,22)(1950,472)
\blacken\path(2040,2362)(1950,2722)(1860,2362)(2040,2362)
\path(1950,2722)(1950,2272)
\blacken\path(6240,2362)(6150,2722)(6060,2362)(6240,2362)
\path(6150,2722)(6150,2272)
\thinlines

\path(1350,2422)(1050,2122)
\path(1350,2422)(1050,2122)
\path(1650,2422)(1050,1822)
\path(1650,2422)(1050,1822)
\path(1950,2422)(1050,1522)
\path(1950,2422)(1050,1522)
\path(1950,2122)(1050,1222)
\path(1950,2122)(1050,1222)
\path(1950,1822)(1050,922)
\path(1950,1822)(1050,922)
\path(1950,1522)(1050,622)
\path(1950,1522)(1050,622)
\path(1950,1222)(1050,322)
\path(1950,1222)(1050,322)
\path(1950,922)(1350,322)
\path(1950,922)(1350,322)
\path(1950,622)(1650,322)
\path(1950,622)(1650,322)
\path(6450,2422)(6150,2122)
\path(6450,2422)(6150,2122)
\path(6750,2422)(6150,1822)
\path(6750,2422)(6150,1822)
\path(7050,2422)(6150,1522)
\path(7050,2422)(6150,1522)
\path(7050,2122)(6150,1222)
\path(7050,2122)(6150,1222)
\path(7050,1822)(6150,922)
\path(7050,1822)(6150,922)
\path(7050,1522)(6150,622)
\path(7050,1522)(6150,622)
\path(7050,1222)(6150,322)
\path(7050,1222)(6150,322)
\path(7050,922)(6450,322)
\path(7050,922)(6450,322)
\path(7050,622)(6750,322)
\path(7050,622)(6750,322)
\thicklines
\path(5550,922)(7650,922)
\path(450,1822)(2550,1822)
\path(450,922)(2550,922)
\path(5550,1822)(7650,1822)
\path(450,1822)(1050,1822)
\blacken\path(690,1732)(1050,1822)(690,1912)(690,1732)
\path(6075,922)(5475,922)
\whiten\path(5835,1012)(5475,922)(5835,832)(5835,1012)
\path(7050,1822)(7650,1822)
\whiten\path(7290,1732)(7650,1822)(7290,1912)(7290,1732)
\path(1950,922)(2550,922)
\whiten\path(2190,832)(2550,922)(2190,1012)(2190,832)
\path(975,922)(375,922)
\whiten\path(735,1012)(375,922)(735,832)(735,1012)
\path(2550,1822)(1950,1822)
\blacken\path(2310,1912)(1950,1822)(2310,1732)(2310,1912)
\path(5550,1822)(6150,1822)
\blacken\path(5790,1732)(6150,1822)(5790,1912)(5790,1732)
\path(7650,922)(7050,922)
\blacken\path(7410,1012)(7050,922)(7410,832)(7410,1012)

\put(-400,1800){$x_{i-1}$}
\put(-400,0900){$x_{i  }$}
\put(4800,1800){$x_{i-1}$}
\put(4800,0900){$x_{i  }$}

\end{picture}

\begin{ca}
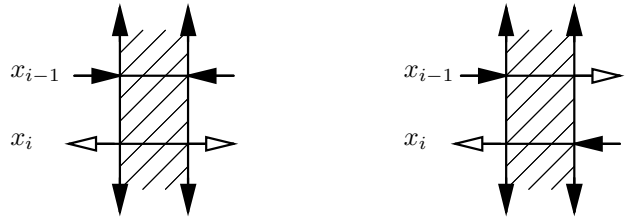
The figure on the left is a schematic presentation of the 
set of configurations with directly opposite inverted 
arrows that we wish to roll up. The figure on the left 
is identical to the first, but the inverted arrow on the 
right is one row higher.
\end{ca}

\end{minipage}
\end{center}

The configuration on the left is the one that we are interested 
in. The configuration on the right is auxiliary. Multiply the 
left configuration by a $b$ vertex, the right configuration 
by a $c$ vertex and consider the sum of the two results 


\begin{center}
\begin{minipage}{4in}
\setlength{\unitlength}{0.001cm}

\begin{picture}(9000,3000)(-1000, 0)

\thicklines

\path(900,2722)(900,22)
\path(1800,2722)(1800,22)
\path(6000,2722)(6000,22)
\path(6900,2722)(6900,22)
\blacken\path(990,2362)(900,2722)(810,2362)(990,2362)
\path(900,2722)(900,2272)
\blacken\path(6990,2362)(6900,2722)(6810,2362)(6990,2362)
\path(6900,2722)(6900,2272)
\blacken\path(810,382)(900,22)(990,382)(810,382)
\path(900,22)(900,472)
\blacken\path(5910,382)(6000,22)(6090,382)(5910,382)
\path(6000,22)(6000,472)
\blacken\path(6810,382)(6900,22)(6990,382)(6810,382)
\path(6900,22)(6900,472)
\blacken\path(1710,382)(1800,22)(1890,382)(1710,382)
\path(1800,22)(1800,472)
\blacken\path(1890,2362)(1800,2722)(1710,2362)(1890,2362)
\path(1800,2722)(1800,2272)
\blacken\path(6090,2362)(6000,2722)(5910,2362)(6090,2362)
\path(6000,2722)(6000,2272)
\thinlines

\path(1200,2422)(900,2122)
\path(1200,2422)(900,2122)
\path(1500,2422)(900,1822)
\path(1500,2422)(900,1822)
\path(1800,2422)(900,1522)
\path(1800,2422)(900,1522)
\path(1800,2122)(900,1222)
\path(1800,2122)(900,1222)
\path(1800,1822)(900,922)
\path(1800,1822)(900,922)
\path(1800,1522)(900,622)
\path(1800,1522)(900,622)
\path(1800,1222)(900,322)
\path(1800,1222)(900,322)
\path(1800,922)(1200,322)
\path(1800,922)(1200,322)
\path(1800,622)(1500,322)
\path(1800,622)(1500,322)
\path(6300,2422)(6000,2122)
\path(6300,2422)(6000,2122)
\path(6600,2422)(6000,1822)
\path(6600,2422)(6000,1822)
\path(6900,2422)(6000,1522)
\path(6900,2422)(6000,1522)
\path(6900,2122)(6000,1222)
\path(6900,2122)(6000,1222)
\path(6900,1822)(6000,922)
\path(6900,1822)(6000,922)
\path(6900,1522)(6000,622)
\path(6900,1522)(6000,622)
\path(6900,1222)(6000,322)
\path(6900,1222)(6000,322)
\path(6900,922)(6300,322)
\path(6900,922)(6300,322)
\path(6900,622)(6600,322)
\path(6900,622)(6600,322)
\thicklines
\path(300,1822)(900,1822)
\blacken\path(540,1732)(900,1822)(540,1912)(540,1732)
\path(5925,922)(5325,922)

\path(825,922)(225,922)

\path(5400,1822)(6000,1822)
\blacken\path(5640,1732)(6000,1822)(5640,1912)(5640,1732)
\path(900,1822)(2100,1822)
\path(900,922)(2100,922)
\path(6000,1822)(7200,1822)
\path(6000,922)(7200,922)

\path(2550,1372)(2100,922)
\blacken\path(2418.198,1631.081)(2100,1822)(2290.919,1503.802)(2418.198,1631.081)
\path(2100,1822)(2550,1372)
\blacken\path(2868.198,1181.081)(2550,1372)(2740.919,1053.802)(2868.198,1181.081)
\path(2550,1372)(3000,922)

\path(3000,1822)(2550,1372)
\thinlines
\path(7200,1822)(8100,922)
\path(8100,1822)(7200,922)
\thicklines

\path(8100,1822)(7650,1372)
\blacken\path(7968.198,1181.081)(7650,1372)(7840.919,1053.802)(7968.198,1181.081)
\path(7650,1372)(8100,922)
\path(7650,1372)(7200,922)
\blacken\path(7390.919,1240.198)(7200,922)(7518.198,1112.919)(7390.919,1240.198)
\path(7200,1822)(7650,1372)

\put(3900,1350){$+      $}

\put(-400,1800){$x_{i-1}$}
\put(-400,0900){$x_{i  }$}

\put(3200,1800){$x_{i  }$}
\put(3200,0900){$x_{i-1}$}

\put(4600,1800){$x_{i-1}$}
\put(4600,0900){$x_{i  }$}

\put(8400,1800){$x_{i  }$}
\put(8400,0900){$x_{i-1}$}

\whiten\path(5685,1012)(5325,922)(5685,832)(5685,1012)
\whiten\path(585,1012)(225,922)(585,832)(585,1012)
\whiten\path(2359.081,1053.802)(2550,1372)(2231.802,1181.081)(2359.081,1053.802)
\whiten\path(2809.081,1503.802)(3000,1822)(2681.802,1631.081)(2809.081,1503.802)
\whiten\path(7909.081,1503.802)(8100,1822)(7781.802,1631.081)(7909.081,1503.802)
\whiten\path(7331.802,1562.919)(7650,1372)(7459.081,1690.198)(7331.802,1562.919)

\end{picture}

\begin{ca}
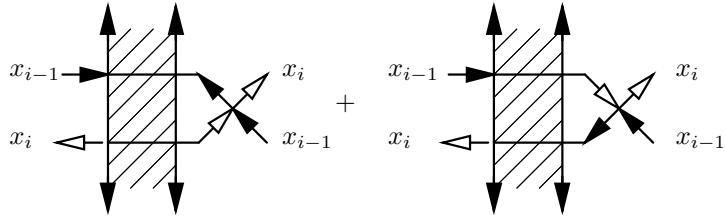
Multiplying the graph on the left above by a type $b$ vertex, and 
that on the right by a type $c$ vertex, and adding the results.
\end{ca}

\end{minipage}
\end{center}

In the sum, all internal arrows, in the external loop on the
right, are now summed over, and the the sum can be represented 
in terms of the single graph on the left of equation 
\ref{anotherequation} 
Applying the Yang Baxter equation to weave that external vertex 
through the lattice we obtain the graph on the right of 
figure \ref{anotherequation} 


\begin{center}
\begin{minipage}{4in}
\setlength{\unitlength}{0.001cm}

\begin{picture}(8500,3000)(-1000, 0)

\thicklines
\path(1050,2722)(1050,22)
\path(1950,2722)(1950,22)
\blacken\path(1140,2362)(1050,2722)(960,2362)(1140,2362)
\path(1050,2722)(1050,2272)
\blacken\path(960,382)(1050,22)(1140,382)(960,382)
\path(1050,22)(1050,472)
\blacken\path(1860,382)(1950,22)(2040,382)(1860,382)
\path(1950,22)(1950,472)
\blacken\path(2040,2362)(1950,2722)(1860,2362)(2040,2362)
\path(1950,2722)(1950,2272)
\thinlines
\path(1350,2422)(1050,2122)
\path(1350,2422)(1050,2122)
\path(1650,2422)(1050,1822)
\path(1650,2422)(1050,1822)
\path(1950,2422)(1050,1522)
\path(1950,2422)(1050,1522)
\path(1950,2122)(1050,1222)
\path(1950,2122)(1050,1222)
\path(1950,1822)(1050,922)
\path(1950,1822)(1050,922)
\path(1950,1522)(1050,622)
\path(1950,1522)(1050,622)
\path(1950,1222)(1050,322)
\path(1950,1222)(1050,322)
\path(1950,922)(1350,322)
\path(1950,922)(1350,322)
\path(1950,622)(1650,322)
\path(1950,622)(1650,322)
\thicklines
\path(450,1822)(1050,1822)
\blacken\path(690,1732)(1050,1822)(690,1912)(690,1732)
\path(975,922)(375,922)

\path(6450,2722)(6450,22)
\path(7350,2722)(7350,22)
\blacken\path(6540,2362)(6450,2722)(6360,2362)(6540,2362)
\path(6450,2722)(6450,2272)
\blacken\path(7440,2362)(7350,2722)(7260,2362)(7440,2362)
\path(7350,2722)(7350,2272)
\blacken\path(6360,382)(6450,22)(6540,382)(6360,382)
\path(6450,22)(6450,472)
\blacken\path(7260,382)(7350,22)(7440,382)(7260,382)
\path(7350,22)(7350,472)
\blacken\path(7635,1012)(7275,922)(7635,832)(7635,1012)
\path(7275,922)(7725,922)

\path(7950,1822)(7500,1822)
\thinlines
\path(6750,2422)(6450,2122)(6450,2047)
\path(7050,2422)(6450,1822)
\path(7350,2422)(6450,1522)
\path(7350,2122)(6450,1222)
\path(7350,1822)(6450,922)
\path(7350,1522)(6450,622)
\path(7350,1222)(6450,322)
\path(7350,922)(6750,322)
\path(7350,622)(7125,322)
\thicklines
\path(1050,1822)(2250,1822)
\path(1050,922)(2250,922)
\path(6150,1822)(7500,1822)
\path(6150,922)(7950,922)
\path(2250,1822)(2700,1372)

\path(3150,1822)(2250,922)
\blacken\path(3018.198,1181.081)(2700,1372)(2890.919,1053.802)(3018.198,1181.081)
\path(2700,1372)(3150,922)
\path(5250,1822)(6150,922)
\blacken\path(5568.198,1631.081)(5250,1822)(5440.919,1503.802)(5568.198,1631.081)
\path(5250,1822)(5700,1372)

\path(5250,922)(6150,1822)

\put(4000,1250){$=      $}

\put(-300,1800){$x_{i-1}$}
\put(-300,0900){$x_{i  }$}

\put(3300,1800){$x_{i  }$}
\put(3300,0900){$x_{i-1}$}

\put(4500,1800){$x_{i-1}$}
\put(4500,0900){$x_{i  }$}

\put(8100,1800){$x_{i  }$}
\put(8100,0900){$x_{i-1}$}

\whiten\path(735,1012)(375,922)(735,832)(735,1012)
\whiten\path(7590,1732)(7950,1822)(7590,1912)(7590,1732)
\whiten\path(2959.081,1503.802)(3150,1822)(2831.802,1631.081)(2959.081,1503.802)
\whiten\path(5440.919,1240.198)(5250,922)(5568.198,1112.919)(5440.919,1240.198)

\end{picture}

\begin{ca}
\label{anotherequation}
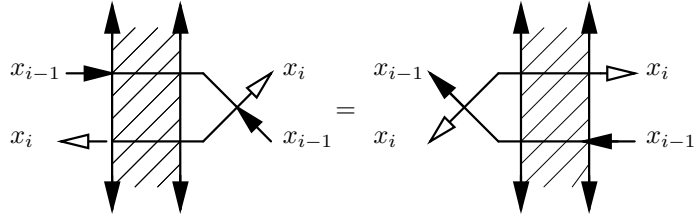
All arrow orientations that are allowed in the external loop on 
the right of the left figure are summed over, so we can use the
Yang Baxter equation to weave that external vertex through the
lattice from the right to the left.
\end{ca}

\end{minipage}
\end{center}

The resulting term on the right hand side of equation 
\ref{anotherequation} can now be decomposed as


\begin{center}
\begin{minipage}{4in}
\setlength{\unitlength}{0.001cm}

\begin{picture}(10000,3000)(-0000, 0)

\thicklines
\path(2100,2722)(2100,22)
\path(3000,2722)(3000,22)
\path(7200,2722)(7200,22)
\path(8100,2722)(8100,22)
\blacken\path(2190,2362)(2100,2722)(2010,2362)(2190,2362)
\path(2100,2722)(2100,2272)
\blacken\path(8190,2362)(8100,2722)(8010,2362)(8190,2362)
\path(8100,2722)(8100,2272)
\blacken\path(2010,382)(2100,22)(2190,382)(2010,382)
\path(2100,22)(2100,472)
\blacken\path(7110,382)(7200,22)(7290,382)(7110,382)
\path(7200,22)(7200,472)
\blacken\path(8010,382)(8100,22)(8190,382)(8010,382)
\path(8100,22)(8100,472)
\blacken\path(2910,382)(3000,22)(3090,382)(2910,382)
\path(3000,22)(3000,472)
\blacken\path(3090,2362)(3000,2722)(2910,2362)(3090,2362)
\path(3000,2722)(3000,2272)
\blacken\path(7290,2362)(7200,2722)(7110,2362)(7290,2362)
\path(7200,2722)(7200,2272)
\thinlines
\texture{55888888 88555555 5522a222 a2555555 55888888 88555555 552a2a2a 2a555555 
	55888888 88555555 55a222a2 22555555 55888888 88555555 552a2a2a 2a555555 
	55888888 88555555 5522a222 a2555555 55888888 88555555 552a2a2a 2a555555 
	55888888 88555555 55a222a2 22555555 55888888 88555555 552a2a2a 2a555555 }
\path(2400,2422)(2100,2122)
\path(2400,2422)(2100,2122)
\path(2700,2422)(2100,1822)
\path(2700,2422)(2100,1822)
\path(3000,2422)(2100,1522)
\path(3000,2422)(2100,1522)
\path(3000,2122)(2100,1222)
\path(3000,2122)(2100,1222)
\path(3000,1822)(2100,922)
\path(3000,1822)(2100,922)
\path(3000,1522)(2100,622)
\path(3000,1522)(2100,622)
\path(3000,1222)(2100,322)
\path(3000,1222)(2100,322)
\path(3000,922)(2400,322)
\path(3000,922)(2400,322)
\path(3000,622)(2700,322)
\path(3000,622)(2700,322)
\path(7500,2422)(7200,2122)
\path(7500,2422)(7200,2122)
\path(7800,2422)(7200,1822)
\path(7800,2422)(7200,1822)
\path(8100,2422)(7200,1522)
\path(8100,2422)(7200,1522)
\path(8100,2122)(7200,1222)
\path(8100,2122)(7200,1222)
\path(8100,1822)(7200,922)
\path(8100,1822)(7200,922)
\path(8100,1522)(7200,622)
\path(8100,1522)(7200,622)
\path(8100,1222)(7200,322)
\path(8100,1222)(7200,322)
\path(8100,922)(7500,322)
\path(8100,922)(7500,322)
\path(8100,622)(7800,322)
\path(8100,622)(7800,322)
\thicklines
\path(8100,1822)(8700,1822)

\path(8700,922)(8100,922)
\blacken\path(8460,1012)(8100,922)(8460,832)(8460,1012)
\path(3075,1822)(3675,1822)

\path(3600,922)(3000,922)
\blacken\path(3360,1012)(3000,922)(3360,832)(3360,1012)
\path(1800,1822)(3000,1822)
\path(1800,922)(3000,922)
\path(6900,1822)(8100,1822)
\path(6900,922)(8100,922)
\path(1800,1822)(900,922)
\path(900,1822)(1800,922)
\blacken\path(1481.802,1112.919)(1800,922)(1609.081,1240.198)(1481.802,1112.919)

\path(900,922)(1350,1372)

\path(1350,1372)(1800,1822)
\path(900,1822)(1350,1372)
\blacken\path(1031.802,1562.919)(1350,1372)(1159.081,1690.198)(1031.802,1562.919)
\path(6900,1822)(6000,922)
\path(6000,1822)(6900,922)
\blacken\path(6709.081,1503.802)(6900,1822)(6581.802,1631.081)(6709.081,1503.802)
\path(6900,1822)(6450,1372)
\path(6000,1822)(6450,1372)
\blacken\path(6131.802,1562.919)(6450,1372)(6259.081,1690.198)(6131.802,1562.919)
\path(6900,922)(6450,1372)

\path(6037,960)(6487,1410)

\put(4500,1372){$+$}
\put(3900,1822){$x_{i  }$}
\put(3900,0922){$x_{i-1}$}
\put(9000,1822){$x_{i  }$}
\put(9000,0922){$x_{i-1}$}
\put(0000,1822){$x_{i-1}$}
\put(0000,0922){$x_{i  }$}
\put(5100,1822){$x_{i-1}$}
\put(5025,0922){$x_{i  }$}

\whiten\path(8340,1732)(8700,1822)(8340,1912)(8340,1732)
\whiten\path(3315,1732)(3675,1822)(3315,1912)(3315,1732)
\whiten\path(1090.919,1240.198)(900,922)(1218.198,1112.919)(1090.919,1240.198)
\whiten\path(1540.919,1690.198)(1350,1372)(1668.198,1562.919)(1540.919,1690.198)
\whiten\path(6768.198,1181.081)(6450,1372)(6640.919,1053.802)(6768.198,1181.081)
\whiten\path(6227.919,1278.198)(6037,960)(6355.198,1150.919)(6227.919,1278.198)
\end{picture}

\begin{ca}
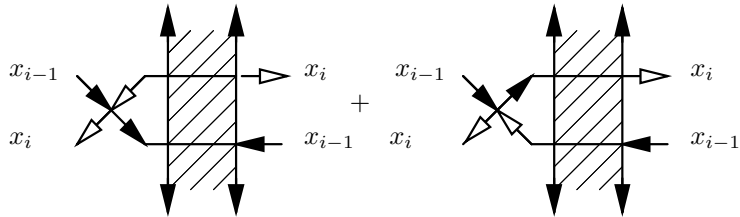
Showing the content of the arrow configurations of the external 
loop on the left.
\end{ca}

\end{minipage}
\end{center}

Dividing all terms by $b(x_{i}, x_{i-1})$, moving a term from the 
left hand side to the right, and using the antisymmetry of the
$b$ weights, we obtain

\begin{equation}
\label{anotheridentityequation}
\ll
{}_{i}^{i}F_{i}^{i} 
\rr
= 
\ll {}_{i-1}^{i  }F_{i-1}^{i  } \rr
      + g_{i, i-1} 
\ll {}_{i  }^{i-1}F_{i  }^{i-1} \rr
      + g_{i-1, i} 
\ll {}_{i  }^{i-1}F_{i-1}^{i  } \rr
\end{equation}

Equation \ref{anotheridentityequation} says that the problem with 
two opposing inverted arrow at row $i$ can be reduced to the same 
problem at row $(i-1)$, plus two boundary 2-point functions of the 
type discussed in Case 3 above. 

Using equation \ref{anotheridentityequation} repeatedly, boundary 
2-point functions with opposite inverted arrows can be computed in 
terms of a sum boundary 2-point functions of type 3.

\begin{re} 
One can show by direct calculation that ${}_{1}^{2}F_{1}^{2}$ 
vanishes, and hence
$$
\ll {}_{2}^{2}F_{2}^{2} \rr = 
        g_{2, 1} 
\ll {}_{2}^{1}F_{2}^{1} \rr
      + g_{1, 2} 
\ll {}_{2}^{1}F_{1}^{2} \rr
$$
\end{re}

\subsection{Boundary $n$-point functions}

\paragraph*{Simple boundary $n$-point functions} can be
constructed almost mechanically by repeated application of 
the above arguments. All that is needed is to develop 
conventions to handle the notational complexity of the 
resulting expressions.

\paragraph*{Composite boundary $n$-point functions} will 
require summing over those inverted boundary arrows that are 
not frozen, which signals a proliferation of terms. However, 
there is hope that such summations can be simplified. 

In \cite{BPZ}, Bogoliubov {\it et al} evaluate not only boundary 
1-point functions, but also {\it the boundary spontaneous
polarizations} which are basically sums over many 1-point 
functions. Bogoliubov {\it et al} use the algebraic Bethe 
{\it ansatz} to obtain expressions that do not contain more terms 
than a typical 1-point function evaluated at the same point 
as the boundary spontaneous polarization.

This leads us to believe that sums that arise in computing 
`composite' boundary $n$-point functions can also be performed, 
leading to relatively simple expressions.

\subsection{Bulk correlation functions}

It is clear how the bulk functions can be obtained as sums 
over products of boundary functions. In the simplest case of 
a 1-point function of a frozen arrow located deep in the lattice, 
the result is a product of two composite boundary correlation 
functions. 

It is highly unlikely that such (typically large) sums over 
products can be used to deduce physical properties of correlation 
functions. To get manageable results, we need to develop efficient 
methods to evaluate the composite boundary functions, which brings 
us back to the comments made in the above paragraphs.

\section{Epilogue}

The point of this work was to show that the result of Bogoliubov 
{\it et al} can reproduced using simple arguments. We also wished
to show the same arguments apply, basically with no modification 
to higher simple boundary functions. The expressions that we 
obtain become cumbersome for higher $n$-functions, but they can 
be written basically by inspection of corresponding graph. 

We hope that there are ways to simplify these expressions by 
performing suitable summations.

\section*{Acknowledgements}

O F wishes to thank J de Gier, M Jimbo and N Kitanine for 
discussions on this work and related topics. I P wishes 
to thank Mr Michael Watt for generously sponsoring his
scholarship.

This research was supported by the Australian Research Council 
(ARC), and partially carried out while I P was visiting the 
Department of Mathematics and Statistics, The University of 
Melbourne.

\end{document}